\documentclass[amsmath,amssymb,aps,prd,superscriptaddress,nofootinbib,onecolumn,11pt]{revtex4-1}

\usepackage{graphicx}
\usepackage{dcolumn}
\usepackage{bm}
\usepackage{hyperref}
\usepackage{color}
\usepackage[utf8]{inputenc}

\begin{document}

\title{Lattice fermion formulation via Physics-Informed Neural Networks: Ginsparg-Wilson relation and Overlap fermions}

\author{Tatsuhiro Misumi}
\email{misumi@phys.kindai.ac.jp}
\affiliation{Department of Physics, Kindai University, Higashi-Osaka, Osaka 577-8502, Japan}

\begin{abstract}
We propose a novel, machine-learning-based framework for constructing lattice fermions using Physics-Informed Neural Networks (PINNs). Our approach treats the formulation of the Dirac operator as an optimization problem guided by physical requirements, such as symmetries, locality and doubler-decoupling conditions. We first demonstrate that, when trained to satisfy the Ginsparg-Wilson (GW) relation as a soft constraint, a neural network reproduces the overlap fermion operator to high numerical accuracy and learns an effective sign-function mapping without explicitly using a prescribed polynomial or rational approximation. 
Secondly, we extend the framework from operator construction to machine-assisted algebraic discovery. Within a generalized polynomial ansatz, the network autonomously drives higher-order terms to zero and recovers the standard Ginsparg-Wilson relation. Remarkably, by changing the initial search bias, the same framework also finds a distinct solution corresponding to a Fujikawa-type generalized GW relation. 
\end{abstract}

\maketitle

\tableofcontents

\newpage

\section{\label{sec:intro}Introduction}

Quantum Chromodynamics (QCD) formulated on a discretized spacetime lattice represents the most robust non-perturbative framework for understanding the strong interaction. However, formulating fermions on a discrete lattice is a long-standing fundamental challenge, primarily due to the fermion doubling problem and its inextricable link to the chiral anomaly.

The fundamental obstacle is rigorously formalized by the Nielsen-Ninomiya no-go theorem \cite{Karsten:1980wd, Nielsen:1980rz, Nielsen:1981xu, Nielsen:1981hk}, which dictates that no single-flavor lattice Dirac operator on the Euclidean spacetime can simultaneously preserve basic conditions: (1) locality, (2) translational invariance, (3) ($\gamma_5$-)hermiticity, and (4) onsite continuous chiral symmetry $\{D, \gamma_5\} = 0$. 

Historically, navigating these stringent constraints has relied exclusively on human analytical ingenuity, effectively creating a game of choosing which fundamental condition to sacrifice. Wilson fermions \cite{Wilson:1975id} successfully decouple the unphysical doublers by introducing an irrelevant dimension-five operator that lifts their masses to the cutoff scale, but at the heavy cost of explicitly breaking chiral symmetry. Staggered fermions \cite{Kogut:1974ag, Susskind:1976jm} distribute spinor components across lattice sites to maintain a remnant symmetry, sacrificing the strict single-flavor interpretation.

A profound paradigm shift occurred with the discovery of the Ginsparg-Wilson (GW) relation \cite{Ginsparg:1981bj}. The GW relation evades the Nielsen-Ninomiya theorem by replacing the exact, ultralocal onsite chiral symmetry with a lattice-deformed version, which was later proved to leave the path integral invariant under a well-defined lattice chiral transformation \cite{Luscher:1998pqa}. This elegant algebraic constraint culminated in exact analytical solutions, most notably the overlap fermion \cite{Neuberger:1998wv} and the closely related domain-wall fermion \cite{Kaplan:1992bt, Shamir:1993zy, Furman:1994ky}.

Despite their theoretical beauty, the exact chiral lattice fermions entail immense practical difficulties. The overlap operator requires the evaluation of a matrix sign function. In standard numerical simulations, this sign function must be approximated via Chebyshev polynomials or Zolotarev rational approximations \cite{Chiu:2002eh,Chiu:2002ir}, which incurs massive computational overhead and introduces truncation errors. Furthermore, the standard overlap fermion often suffers from delicate real-space locality issues that depend heavily on the smoothness of the background gauge fields. To circumvent these bottlenecks, lattice theorists have spent decades proposing improved versions, such as optimized and approximate GW operators \cite{Gattringer:2000js,Bietenholz:1998ut, Bietenholz:1999km}, Brillouin-improved fermions \cite{Durr:2010ch,Durr:2012dw, Cho:2013yha,Cho:2015ffa,Durr:2017wfi}, and related improved kernel constructions \cite{Creutz:2010bm,Misumi:2012eh},
which utilize hypercubic isotropic derivatives to drastically optimize the dispersion relation and improve locality.

Concurrently, machine learning (ML) techniques have revolutionized numerous scientific domains. Particularly, Physics-Informed Neural Networks (PINNs) \cite{Raissi:2017zsi} have proven that deep learning models can be constrained by fundamental physical laws---such as partial differential equations or algebraic symmetries---embedded directly into their loss functions. While ML has seen rapid adoption in lattice QCD for configuration generation \cite{Boyda:2022nmh,Tomiya:2025quf,Wenger:2026sjp}, solving inverse problems and the formulation of improved gauge actions \cite{Holland:2023ews,Holland:2024muu,Wenger:2025sre,Holland:2025fsa}, its capacity as an engine for the fundamental formulation of the Dirac action itself remains an uncharted frontier \cite{Yasunaga:2026smj}.

In this paper, we propose a fundamental paradigm shift: utilizing PINNs not only to construct lattice fermion operators but to discover their underlying algebraic structures. Specifically, in the free-field setting, we demonstrate that a generic neural network trained with the GW relation as a soft constraint can reproduce the overlap Dirac operator to high numerical accuracy. 
Furthermore, by implementing a differentiable architecture search (DARTS \cite{Liu:2018alq})-inspired optimization over a generalized polynomial ansatz, we demonstrate that the network can autonomously recover the standard Ginsparg-Wilson relation from an agnostic initial condition. This recovery should be understood within the prescribed polynomial search space: starting from unbiased coefficients, the optimization selects the $n=1$ structure and suppresses higher-order terms under the combined requirements of locality and doubler decoupling. Strikingly, by changing the initial logits, the same framework also finds a distinct solution corresponding to a Fujikawa-type generalized GW relation \cite{Fujikawa:2000my,Clancy:2023ino}.

The idea of optimizing lattice discretizations under physical constraints has a long history in the program of perfect and fixed-point  actions~\cite{Hasenfratz:1993sp,Bietenholz:1995cy}.  Optimized Dirac operators, including overlap-hypercube constructions, further showed that improved locality and dispersion properties can be pursued together with chiral symmetry constraints~\cite{Bietenholz:2006ni}. Our PINN framework can be interpreted as providing a complementary machine-learning route to such an optimization problem.

This paper is organized as follows. In Section \ref{sec:framework}, we outline the general methodology of our machine-learning framework and its potential extensions. In Section \ref{sec:theory}, we review the foundational mechanics of the GW relation and its relation to overlap fermions. In Section \ref{sec:overlap_discovery}, we demonstrate that a neural network can successfully learn the overlap mapping. 
We also discuss its straightforward application to 4D. 
Finally, in Section \ref{sec:autonomous_algebra}, we present the main result of this study: the autonomous algebraic discovery of the GW relation and its higher-order generalizations. Conclusions and future directions are summarized in Section \ref{sec:conclusion}.
To demonstrate the stability of our framework, we provide an analysis of its robustness against variations in the loss weights in Appendix \ref{app:weights}, and its sensitivity to the locality decay parameter $\alpha$ in Appendix \ref{app:alpha}.

\section{\label{sec:framework}Lattice Fermions from Soft Constraints}

The core philosophy of our proposal is to replace the analytical derivation of lattice operators with constraint-driven machine learning. A generic, translationally invariant lattice Dirac operator $D(k)$ can be parameterized by a highly expressive function approximator, such as a Multi-Layer Perceptron (MLP) defined in momentum space.

The desired physical properties of the fermion are translated into ``soft constraints'' within a total loss function during the training phase,
\begin{equation}
    \mathcal{L}_{total} \,=\, \sum_i w_i \mathcal{L}_i(D)\,.
\end{equation}
The constraints, or loss functions, can naturally include:
\begin{itemize}
    \item \textbf{Symmetry constraints ($\mathcal{L}_{sym}$)}: Enforcing continuous symmetries, discrete symmetries, or deformed algebraic structures.
    \item \textbf{Locality constraints ($\mathcal{L}_{loc}$)}: Penalizing long-range real-space hoppings via Fourier transforms to encourage physically acceptable locality properties.
    \item \textbf{Spectrum constraints ($\mathcal{L}_{spec}$)}: Explicitly suppressing doublers, or pinning doubler locations in momentum space.
\end{itemize}

By carefully balancing the relative weights $w_i$, called hyperparameters, the neural network systematically navigates the landscape of valid lattice actions. The model discovers optimal compromises between competing physical requirements (e.g., extreme locality versus exact symmetry), effectively automating the design of lattice field theories.

This modular framework provides immense flexibility and is strictly not limited to the standard single-flavor Dirac operator. For instance, the methodology can be naturally extended to one-component fermions such as staggered fermions. Furthermore, it opens new robust avenues for exploring minimally doubled fermions, where specific discrete symmetries ($\mathcal{C, P, T}$) or full hypercubic symmetries are intentionally broken \cite{Karsten:1981gd,Wilczek:1987kw,Creutz:2007af,Borici:2007kz,Bedaque:2008xs,Bedaque:2008jm, Capitani:2009yn,Kimura:2009qe,Kimura:2009di,Creutz:2010cz,Creutz:2010qm,Capitani:2010nn,Tiburzi:2010bm,Kamata:2011jn,Misumi:2012uu,Misumi:2012ky,Capitani:2013zta,Capitani:2013iha,Misumi:2013maa,Weber:2013tfa,Weber:2017eds,Durr:2020yqa}. The framework can also be adapted to the Hamiltonian formalism of lattice fermions, establishing a solid foundation for future studies aiming at quantum simulations and tensor network applications \cite{Li:2024dpq,Chatterjee:2024gje,Catterall:2025vrx,Yamaoka:2025sdm,Onogi:2025xir,Aoki:2025vtp,Seiberg:2026icc,Misumi:2025yjf,Misumi:2026ckr}.

\section{\label{sec:theory}Review of the Ginsparg-Wilson relation and Overlap fermions}

To provide a concrete realization of our PINN framework, we target the construction of a GW-compliant fermion. We briefly review the underlying mechanics of lattice chiral symmetry to formalize our optimization objectives.

In continuum Euclidean space, the massless Dirac operator $D = \gamma_\mu \partial_\mu$ perfectly anti-commutes with $\gamma_5$. A naive symmetric finite-difference discretization yields the momentum-space operator
\begin{equation}
    D_{naive}(k) \,=\, \frac{i}{a} \sum_\mu \gamma_\mu \sin(a k_\mu)\,.
    \label{eq:naive}
\end{equation}
While $\{ \gamma_5, D_{naive} \} = 0$ holds, $D_{naive}(k)$ possesses zeros at $2^d$ points in the Brillouin zone, corresponding to $2^d$ doublers.
From now, we consider the dimensionless operators and momenta as $a D \to D$ and $a k \to k$.

This is resolved by adding a momentum-dependent mass term, or the Wilson kernel $M_W$,
\begin{eqnarray}
    D_W(k) &=& i \sum_\mu \gamma_\mu \sin k_\mu + M_W(k)\,, \nonumber \\
    M_W(k) &=& m_0 + r \sum_\mu (1 - \cos k_\mu)\,.
    \label{eq:wilson}
\end{eqnarray}
The physical mode at $k=0$ retains the mass $m_0$, whereas the doublers acquire masses of $O(1/a)$. The unavoidable cost is the explicit breaking of chiral symmetry $\{ \gamma_5, D_W \}  \neq 0$.

It was proposed \cite{Ginsparg:1981bj} that the anti-commutation relation should be deformed by a lattice artifact as
\begin{equation}
    D\gamma_5 \,+\, \gamma_5 D \,=\, D \gamma_5 D\,.
    \label{eq:gw}
\end{equation}
Assuming $\gamma_5$-hermiticity $D^\dagger = \gamma_5 D \gamma_5$, multiplying Eq.~(\ref{eq:gw}) by $\gamma_5$ gives $D + D^\dagger = D^\dagger D$. Thus, the eigenvalues of any operator satisfying the GW relation must lie exactly on a circle in the complex plane, centered at $(1, 0)$ with radius $1$.
An exact analytical solution to this relation is the overlap Dirac operator \cite{Neuberger:1998wv}
\begin{equation}
    D_{ov} \,=\,  1 \,+\, \gamma_5\, \text{sgn}(\gamma_5 D_W)\, .
    \label{eq:overlap_op}
\end{equation}

A crucial aspect of exact chiral lattice fermions is their spatial locality. Unlike the ultralocal Wilson Dirac operator, operators satisfying the GW relation, such as the overlap fermion, are inherently non-ultralocal (i.e., their couplings do not strictly vanish beyond a finite distance). However, it has been rigorously shown that as long as the underlying input kernel is sufficiently smooth and local, the resulting overlap operator exhibits strict exponential decay in real space $|D(x,y)| \le C e^{-\alpha \|x-y\|}$ \cite{Hernandez:1998et,Horvath:1998cm,Bietenholz:1999dg}. Consequently, explicitly incorporating and evaluating the exponential spatial locality of the derived operator---characterized by the decay rate $\alpha$---is a paramount criterion when exploring the landscape of GW-compliant solutions via machine learning.

\section{\label{sec:overlap_discovery} Deep Learning of Overlap Operator}

We first demonstrate that a neural network can successfully learn the exact overlap mapping to high numerical accuracy when the GW relation is explicitly provided as a target constraint. 

We parameterize the Dirac operator using an MLP defined in momentum space. To ensure the network learns a universal mapping rather than overfitting to a specific configuration, we employ domain randomization by uniformly sampling the bare mass $m_0 \in [-1.75, -0.25]$ at each training epoch. The total loss function is defined as 
\begin{equation}
\mathcal{L}_{total} \,=\, w_{GW} \mathcal{L}_{GW} \,+\, w_{pin} \mathcal{L}_{pin} \,+\, w_{loc}\mathcal{L}_{loc}\,.
\end{equation}
Here, $\mathcal{L}_{GW}$ penalizes deviations from the exact GW relation, defined as 
\begin{align}
\mathcal{L}_{GW}\,=\, \frac{1}{N_k}\sum_{k}||\{D_\theta (k),\gamma_5\} - D_\theta (k)\gamma_5 D_\theta (k)||^2\,.
\end{align}
The detailed roles of $\mathcal{L}_{pin}$ and $\mathcal{L}_{loc}$ are elaborated in the following subsection.

\subsection{Neural Network Implementation Details}

The Dirac operator is modeled by a Multi-Layer Perceptron (MLP) with three hidden layers (256 units each) using smooth $\tanh$ activation functions to ensure infinite differentiability in momentum space. 
For the 2D case, the input layer of the network takes a 3-dimensional vector consisting of the basic lattice momentum functions
\begin{equation}
\text{Input}:\,\, \left( \sin k_1, \sin k_2, M_W(k) \right) \in \mathbb{R}^3 \,.
\end{equation}
The network, parameterized by weights $\theta$, then outputs three real-valued components $\left({\cal V}_{1}, {\cal V}_{2}, {\cal M} \right) \in \mathbb{R}^3$, which construct the general momentum-space Dirac operator as
\begin{equation}
\text{Output}:\,\, D_{\theta}(k) \,=\, i \sum_{\mu=1}^2 \gamma_{\mu} {\cal V}_{\mu}(k) \,+\, \mathbb{I} {\cal M}(k)\,.
\end{equation}
For 4D, the inputs and outputs are straightforwardly expanded to 5 dimensions to accommodate the four Euclidean gamma matrices and mass or Wilson terms.
In essence, our neural network functions as a highly non-linear map that takes the conventional Wilson fermion kernel as its input and yields an optimized chiral lattice fermion operator as its output.

Parity symmetry is explicitly enforced by symmetrizing the network's forward pass: $\mathcal{V}_\mu(k) = \frac{1}{2}(\mathcal{V}_\mu^{fwd}(k) - \mathcal{V}_\mu^{bwd}(-k))$ and $\mathcal{M}(k) = \frac{1}{2}(\mathcal{M}^{fwd}(k) + \mathcal{M}^{bwd}(-k))$, ensuring that the spatial vector components are exactly odd functions, while the scalar mass component is strictly an even function of momentum.

In our loss function formulation, $\mathcal{L}_{pin}$ plays a critical role in enforcing the doubler-decoupling conditions. This spectrum constraint explicitly requires the physical zero mode to remain massless ($D(0)=0$) while forcefully pushing the unphysical doubler modes to the cutoff scale ($D(k_{doub}) = 2$), leading to the loss function
\begin{align}
\mathcal{L}_{pin} = || D_{\theta}(0,0) ||^2 + || D_{\theta}(0,\pi) -2{\mathbb I}||^2 + || D_{\theta}(\pi,0) -2{\mathbb I}||^2
+ || D_{\theta}(\pi,\pi) -2{\mathbb I}||^2\,. 
\label{eq:pin}
\end{align}
Without this doubler pinning loss, the network might easily fall into trivial parameter spaces, such as a null operator. Alternatively, this strict penalty can be relaxed into an inequality constraint, such as $||D_\theta(k_{doub})-1.5{\mathbb I}|| > 0$, by employing a ReLU-based loss function.

Finally, the locality penalty $L_{\rm loc}$ penalizes long-range real-space hoppings and thereby biases the learned operator toward an exponentially localized profile such as $|D(r)| \sim e^{-\alpha |r|}$ discussed in Sec.~\ref{sec:theory}. This is implemented via a Fast Fourier Transform (FFT) to real space as $D_{\theta}(r) = {\rm FFT}[D_{\theta}(k)]$, applying a weighted exponential penalty 
\begin{align}
\mathcal{L}_{loc} = \sum_r || D_{\theta}(r) \cdot (e^{\alpha |r|} - 1) ||^2 \,. 
\label{eq:loc}
\end{align}
Here, the distance $|r|$ is defined as the shortest Euclidean distance on the periodic lattice. For a $d$-dimensional lattice of linear size $L$, it is explicitly calculated as $|r| = \sqrt{\sum_\mu \min(|x_\mu|, L-|x_\mu|)^2}$, which properly accounts for the periodic boundary conditions.

For the investigation in this section, we utilize the lattice size $L=32$ with a decay parameter $\alpha=0.8$ for 2D, and $L=10$ with $\alpha=0.8$ for 4D. 
The network parameters are optimized using the Adam optimizer. The learning rate is initialized to $\eta = 0.002$ and is smoothly annealed during the training process to ensure stable convergence. The relative weights of the total loss function are chosen as $w_{GW} = 1.0$, $w_{pin} = 0.1$, and $w_{loc} = 0.1$. This specific weighting prevents the harsh spatial locality penalty from overpowering the optimization in the early epochs, allowing the network to first satisfy the exact algebraic and spectrum constraints before fine-tuning the spatial profile. 

During the training process, the network smoothly navigates the complex optimization landscape. While the initial random weights produce arbitrary spectra that violate all symmetries, the simultaneous minimization of these three distinct loss components steadily guides the network toward a GW-compliant operator with suppressed long-range hoppings.

To explicitly demonstrate the robustness of our framework against the choice of these specific hyperparameters, we provide a stability analysis against variations in the loss weights in Appendix \ref{app:weights}, and against variations in the locality decay parameter $\alpha$ in Appendix \ref{app:alpha}.
Furthermore, in addition to the hyperparameter stability discussed above, to verify the robustness and reproducibility of our framework against different network initializations, we performed multiple independent training runs with different random initializations. Across all trials (each with 10,000-20,000 epochs), the network consistently converged to nearly identical physical results, demonstrating the high stability of the optimization process.

\subsection{Results: Learned Overlap Mapping in 2D}

\begin{figure}[t]
\centering
\includegraphics[width=0.8\linewidth]{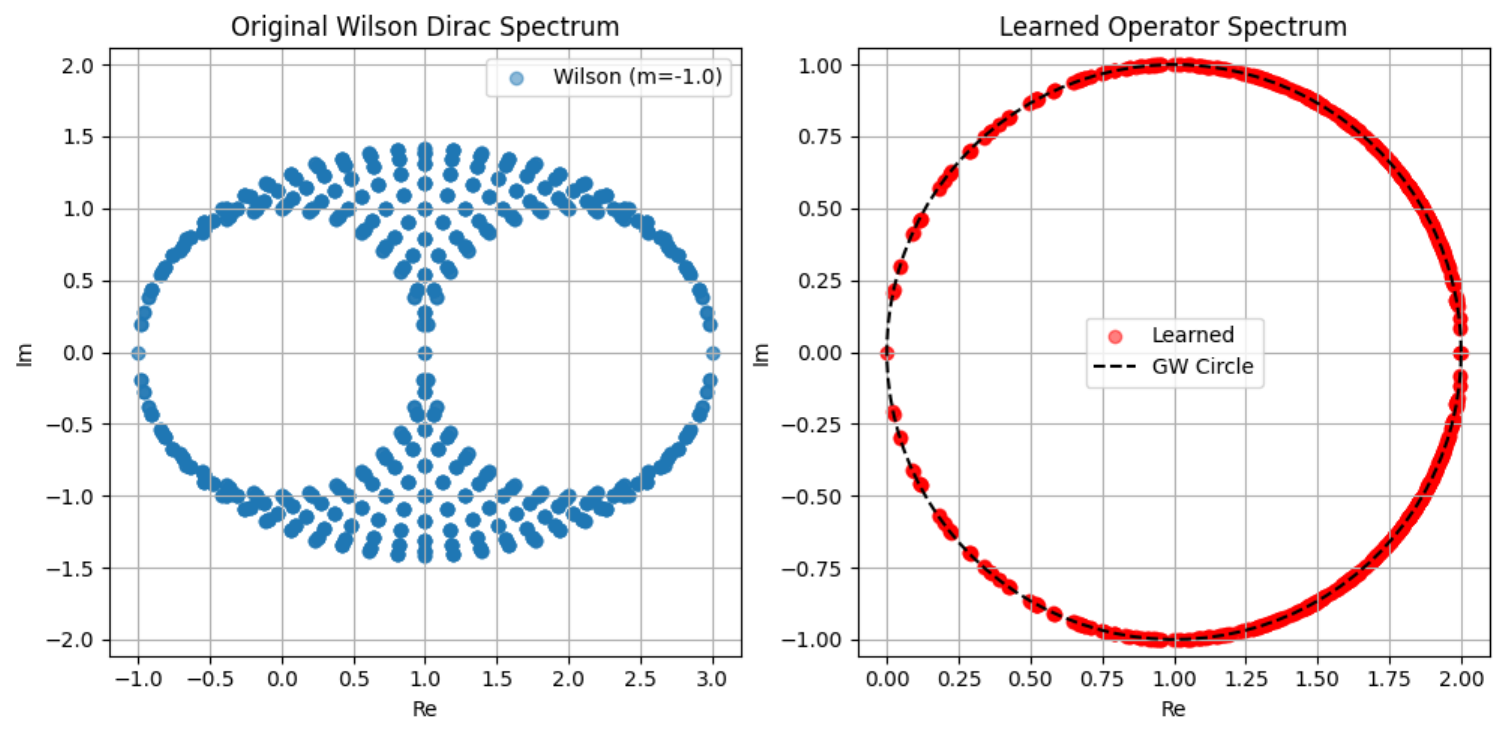}
\caption{Comparison of the eigenvalue spectrum in the complex plane. Left: The original 2D Wilson Dirac spectrum (input). Right: The spectrum of the learned 2D operator (output), which collapses onto the Ginsparg-Wilson circle. While $m_0$ is randomized during training, the spectrum is evaluated here at $m_0 = -1.0$. The parameters are $\alpha=0.8$, $L=32$ and $w_{GW} = 1.0$, $w_{pin} = 0.1$, $w_{loc} = 0.1$.}
\label{fig:spectrum_2d}
\end{figure}

Post-training analysis demonstrates a good agreement with the GW formulation. As shown in Fig.~\ref{fig:spectrum_2d}, the eigenvalue spectrum of the learned operator collapses onto the GW circle. It is notable that, while $m_0$ is randomized during training, the spectrum is evaluated here at $m_0 = -1.0$. The near zero mode deviates from the origin by approximately $3.3870 \times 10^{-4}$, showing the high precision of the doubler-pinning loss. To test the accuracy of the learned mapping, we extract the effective sign function $\epsilon_{eff}$ from the learned operator and project it against the eigenvalues of the original Wilson Hamiltonian $H_W = \gamma_5 D_{W}$. 

\begin{figure}[t]
\centering
\includegraphics[width=0.6\linewidth]{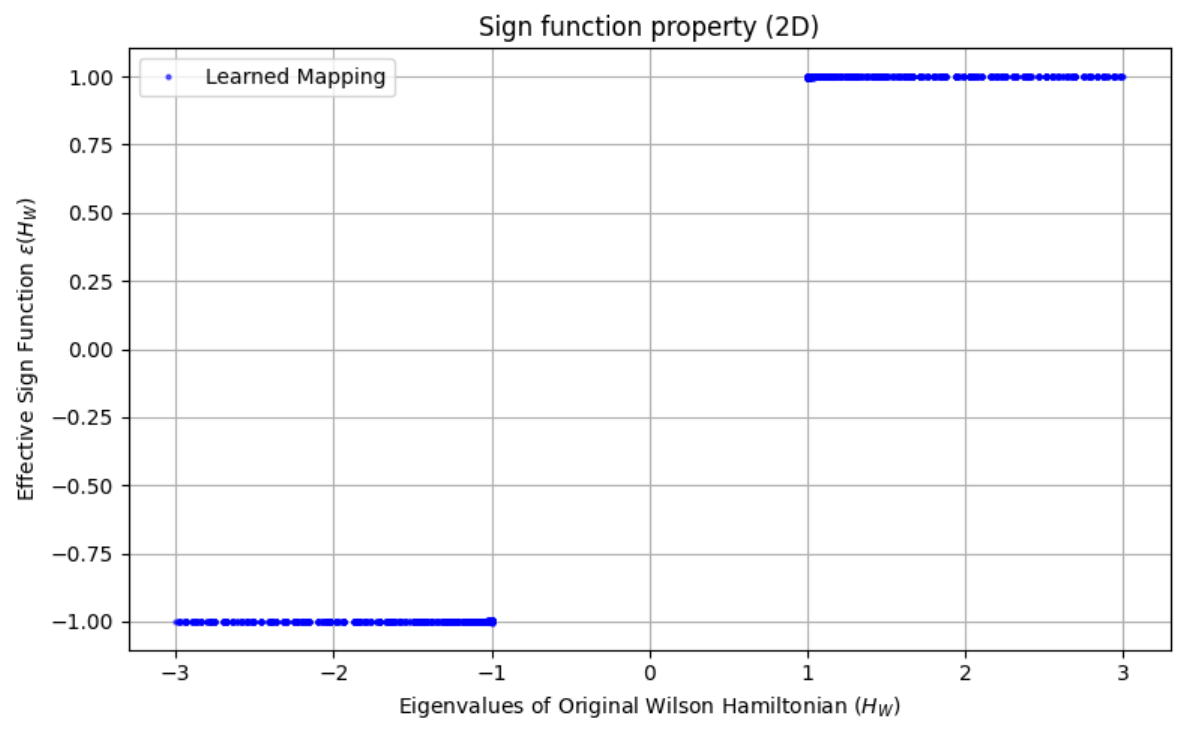}
\caption{The effective sign function $\epsilon_{eff}(H_W)$ extracted from the learned operator, plotted against the eigenvalues of the original Wilson Hamiltonian $H_W$. The network realizes a sharp step function. The parameters are $\alpha=0.8$, $L=32$ and $w_{GW} = 1.0$, $w_{pin} = 0.1$, $w_{loc} = 0.1$.}
\label{fig:sign_func_2d}
\end{figure}

As illustrated in Fig.~\ref{fig:sign_func_2d}, the extracted effective sign function forms a sharp step-like profile and is numerically saturated near $\pm 1$ on the sampled eigenvalues of $H_W$. Quantitatively, the maximum absolute deviation of the learned points from the exact theoretical sign function is $\sim 5.2439\times 10^{-3}$, with a root-mean-square error (RMSE) $\sim 1.0845 \times 10^{-3}$. The result shows that the nonlinear neural-network ansatz can represent the overlap sign-function mapping with high numerical accuracy, without explicitly specifying a polynomial or rational approximation scheme.

\begin{figure}[t]
\centering
\includegraphics[width=0.6\linewidth]{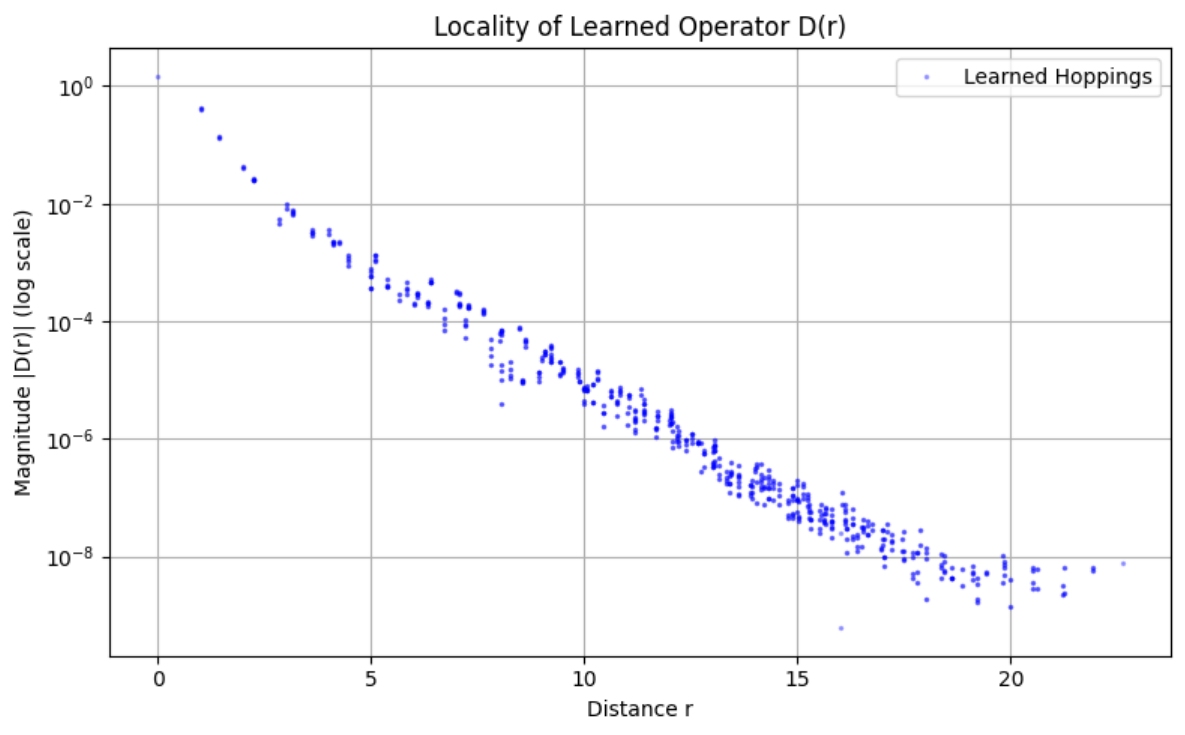}
\caption{Magnitude of the real-space hopping parameters $|D(r)|$ of the learned 2D operator as a function of the distance $r$. The observed exponential-like suppression provides a numerical check of the locality properties of the learned operator. The parameters are $\alpha=0.8$, $L=32$ and $w_{GW} = 1.0$, $w_{pin} = 0.1$, $w_{loc} = 0.1$.}
\label{fig:locality_2d}
\end{figure}

\begin{figure}[t]
\centering
\includegraphics[width=\linewidth]{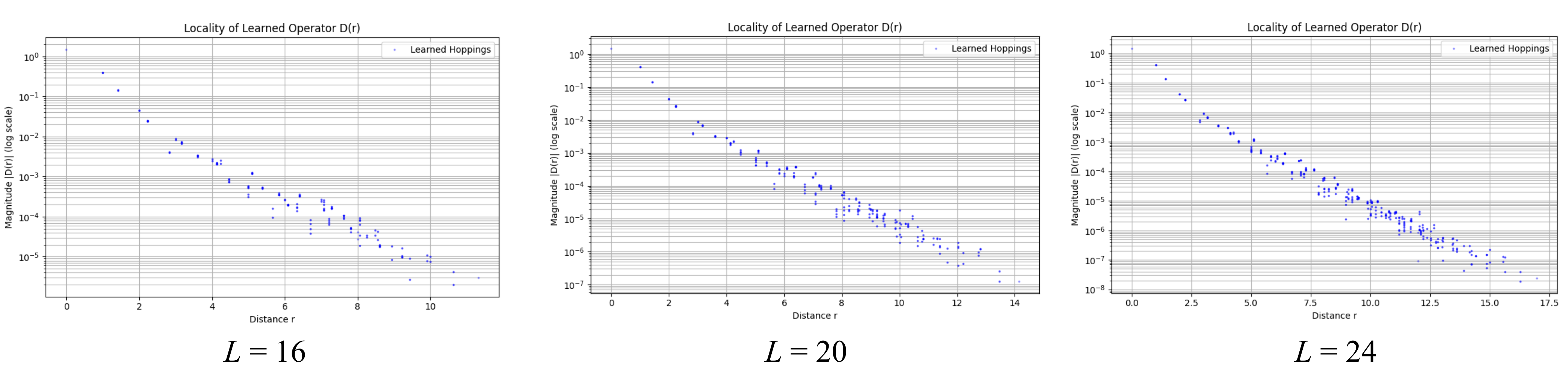}
\caption{Magnitude of the real-space hopping parameters $|D(r)|$ of the learned 2D operator as a function of the distance $r$ for smaller lattices $L=16,20,24$ with $\alpha=0.8$ and $L=32$. The plateau shifts according to the volume.}
\label{fig:finiteV}
\end{figure}

Real-space checks in Fig.~\ref{fig:locality_2d} show that the hoppings of the learned operator are strongly suppressed at large distances, consistently with an exponentially localized profile. 
An exponential fit $C e^{-\mu |r|}$ to the decay behavior in the intermediate range yields a fitted locality exponent of $\mu \sim 1.02$. Furthermore, at very large distances, the locality plot exhibits a plateau around $O(10^{-8})$. To clarify whether this feature is a finite-volume artifact or an immanent artifact of the numerical procedure, we performed identical training on smaller lattice volumes $L=16,20,24$ in Fig.~\ref{fig:finiteV}. The comparison reveals that for these smaller volumes, the plateau does not appear, as the maximum distance is reached before the magnitude hits the $O(10^{-8})$ noise floor. This clearly indicates that the plateau primarily originates from the numerical precision limit (single-precision floating-point format) of the current network training, rather than physical finite volume effects.

\begin{figure}[t]
\centering
\includegraphics[width=\linewidth]{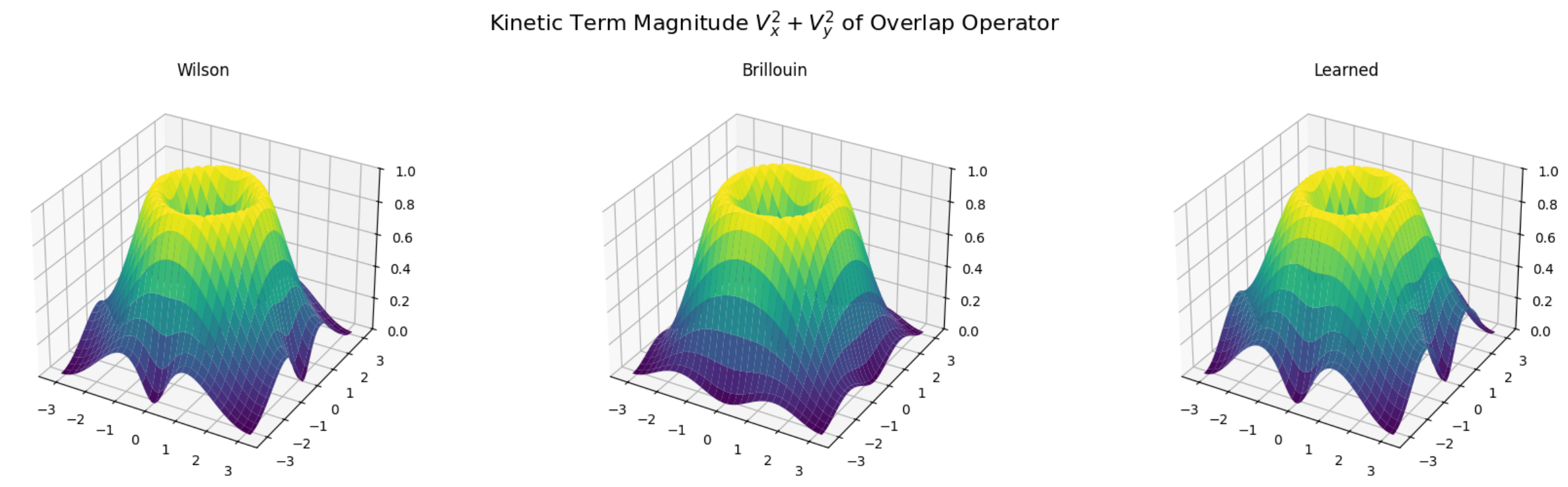}
\caption{Plots of the kinetic term magnitude $|{\mathcal V}_x|^2 + |{\mathcal V}_y|^2$. Left: Standard Wilson-overlap. Center: Brillouin-overlap with characteristic flat doubler basins. Right: The learned operator with $\alpha=0.8$, $L=32$ and $w_{GW} = 1.0$, $w_{pin} = 0.1$, $w_{loc} = 0.1$.}
\label{fig:dispersion_2d-1}
\end{figure}

\begin{figure}[t]
\centering
\includegraphics[width=\linewidth]{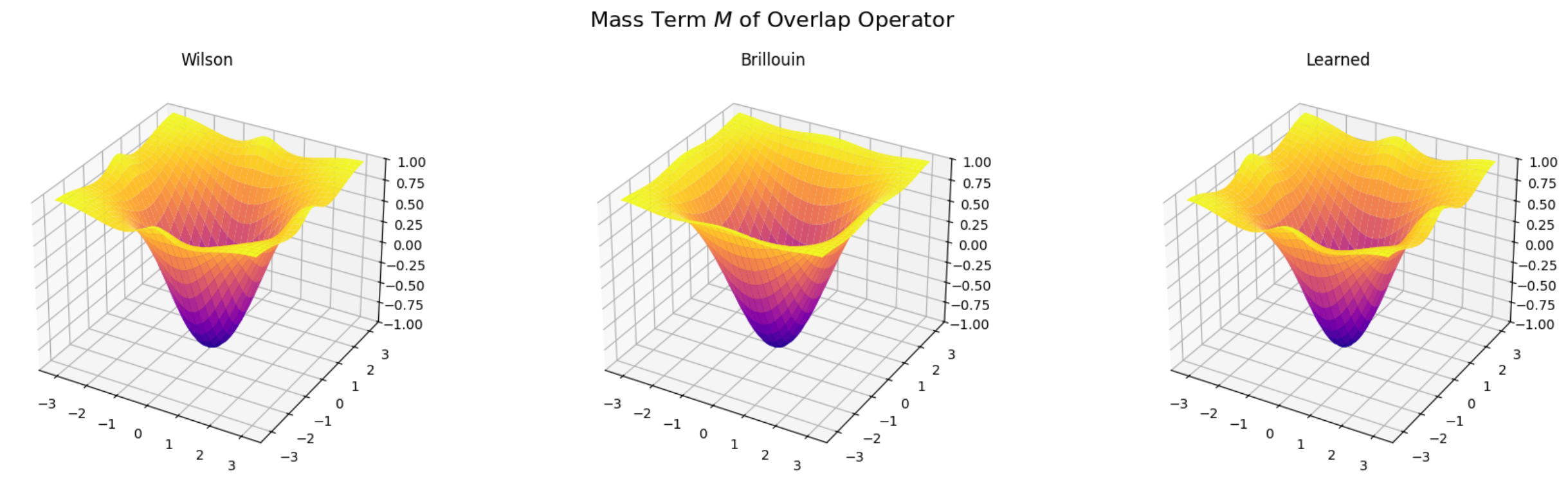}
\caption{Plots of the mass term ${\mathcal M}$. Left: Standard Wilson-overlap. Center: Brillouin-overlap with characteristic flat doubler basins. Right: The learned operator with $\alpha=0.8$, $L=32$ and $w_{GW} = 1.0$, $w_{pin} = 0.1$, $w_{loc} = 0.1$.}
\label{fig:dispersion_2d-2}
\end{figure}

We also plot the kinetic term magnitude $|{\mathcal V}_x(k)|^2 + |{\mathcal V}_y(k)|^2$ in Fig.~\ref{fig:dispersion_2d-1} and the momentum-dependent mass term ${\mathcal M}(k)$ in Fig.~\ref{fig:dispersion_2d-2} for Wilson-overlap, Brillouin-overlap \cite{Durr:2010ch,Durr:2012dw} and our learned operator. By examining the dispersion near the Brillouin zone boundaries, we observe that the operator learned by the neural network closely resembles the standard Wilson-overlap operator rather than the Brillouin-overlap, under the present choice of $w_{loc}$ and $\alpha$. Notably, the neural network framework has the potential to discover improved formulations, such as the Brillouin-overlap operator, through appropriate tuning of the penalty parameters $w_{loc}$ and the decay parameter $\alpha$. We investigate this issue in Appendix \ref{app:weights}.

\subsection{Extension to 4D}

\begin{figure}[t]
\centering
\includegraphics[width=0.8\linewidth]{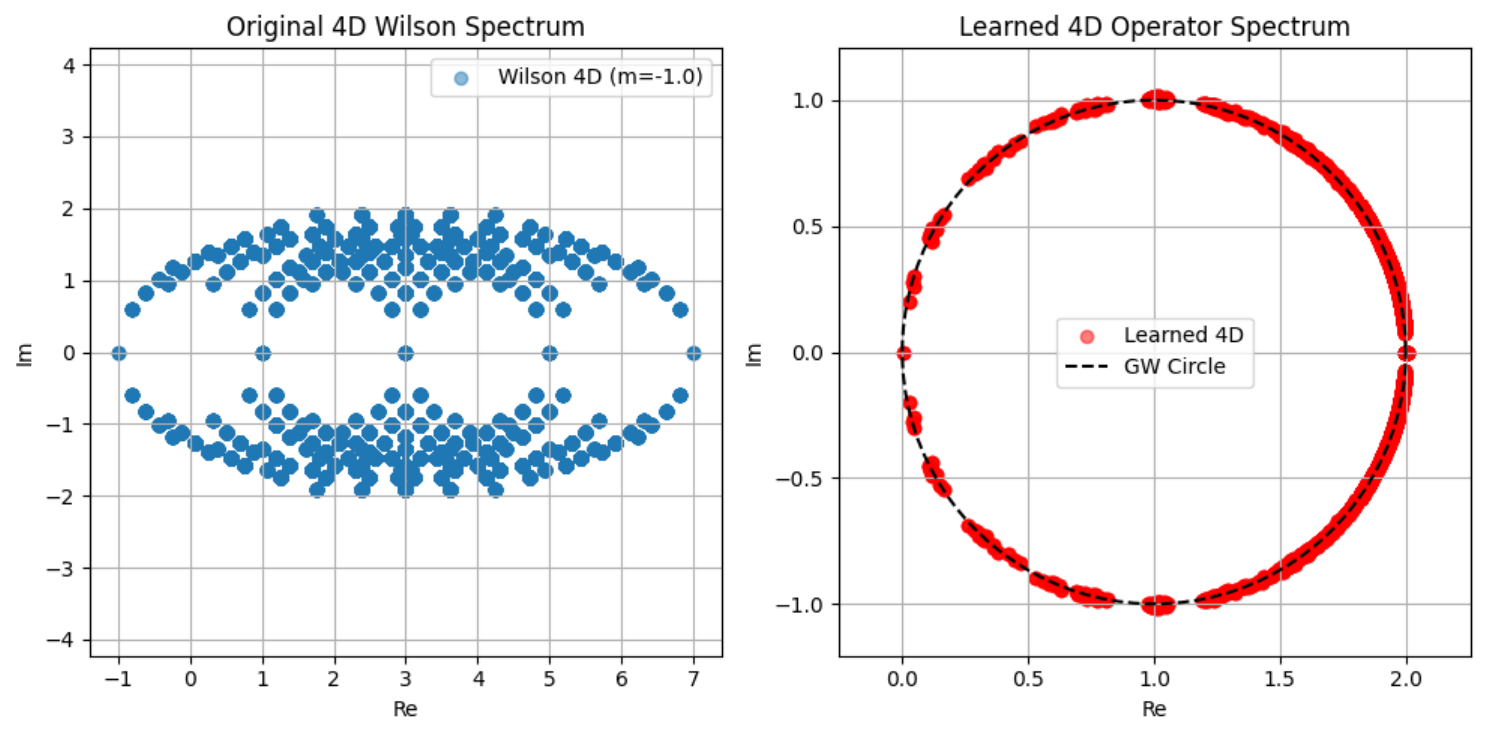}
\caption{The 4D spectrum of the learned operator. Left: The original 4D Wilson Dirac spectrum (input). Right: The spectrum of the learned 4D operator (output), which lies close to the Ginsparg-Wilson circle. While $m_0$ is randomized during training, the spectrum is evaluated here at $m_0 = -1.0$. The parameters are $\alpha=0.8$, $L=10$ and $w_{GW} = 1.0$, $w_{pin} = 0.1$, $w_{loc} = 0.1$.}
\label{fig:4d_spec}
\end{figure}

We successfully extend our methodology to four dimensions to confirm its broad applicability. The network architecture is expanded to output five components, and the exponential locality penalty is evaluated via a 4D FFT over a hypercubic grid. The optimization successfully converges, and the eigenvalue spectrum of the resulting 4D operator lies close to the GW circle as shown in Fig.~\ref{fig:4d_spec}, where we set $w_{GW} = 1.0$, $w_{pin}=0.1$, $w_{loc}=0.1$, $\alpha = 0.8$ and $L=10$. The near zero mode deviates from the exact origin by only $\sim 4.8639 \times 10^{-3}$, demonstrating that the doubler-pinning condition is accurately satisfied in 4D.

\begin{figure}[t]
\centering
\includegraphics[width=0.6\linewidth]{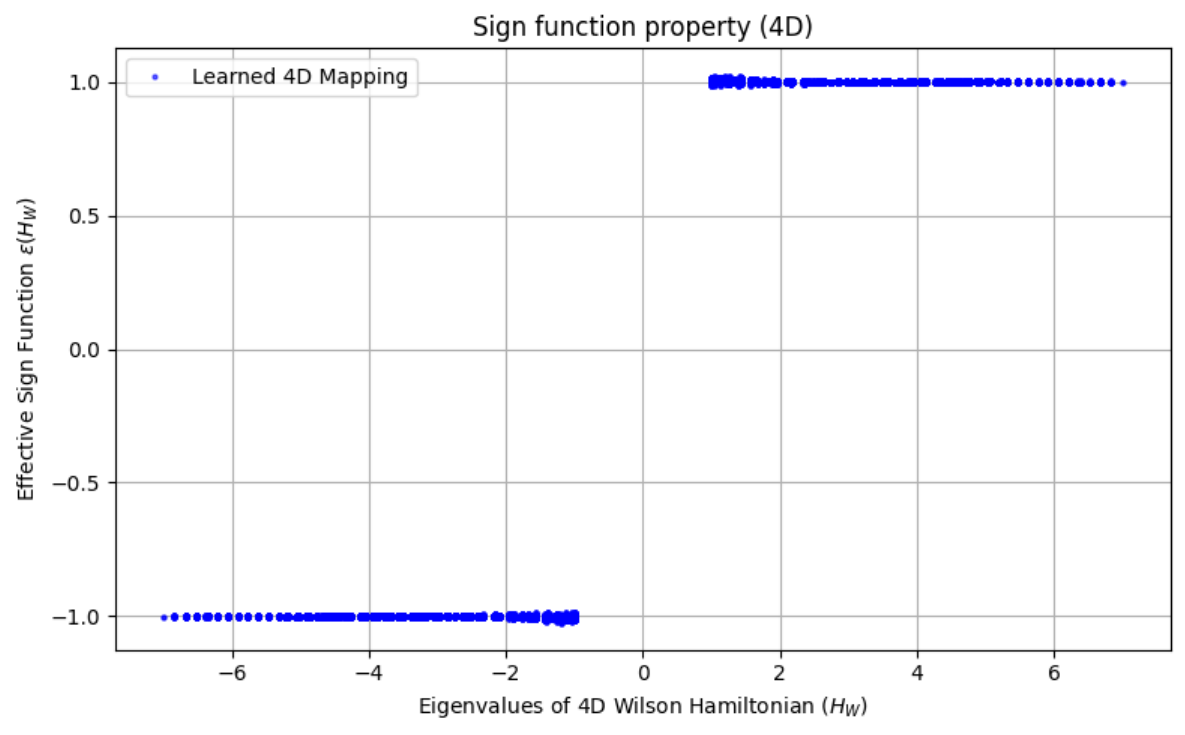}
\caption{
The effective sign function $\epsilon(H_W)$ extracted from the 4D learned operator, where the network realizes a step function. The parameters are $\alpha=0.8$, $L=10$ and $w_{GW} = 1.0$, $w_{pin} = 0.1$, $w_{loc} = 0.1$.}
\label{fig:4d_sign}
\end{figure}

The effective sign function similarly realizes a step-like function as shown in Fig.~\ref{fig:4d_sign}. The maximum absolute deviation from the exact 4D sign function is $\sim 3.1569 \times 10^{-2}$ with RMSE $\sim 1.4993 \times 10^{-3}$. This demonstrates that the machine-learning framework can be similarly applied to higher dimensions.
However, in this 4D study, we only have the small lattice as $L=10$ and need larger lattices to investigate the locality and the property of bare kernels as done for 2D. Future work will be devoted to this investigation.

\section{\label{sec:autonomous_algebra}Autonomous Discovery of Ginsparg-Wilson Relation}

The previous section demonstrated that a neural network can solve the problem of finding a local lattice operator when explicitly given the GW relation ($D+D^\dagger =  D^\dagger D$) as a target constraint. However, a more ambitious question remains: can the network recover the GW algebraic structure within a prescribed polynomial search space, guided by locality and doubler-decoupling constraints?

To answer this, we extend our framework from inverse construction to machine-assisted algebraic-structure search. We abandon the explicit GW constraint $\mathcal{L}_{GW}$ and instead propose a generalized polynomial ansatz for the algebraic structure,
\begin{equation}
    D + D^\dagger \,=\, \sum_{n=1}^{N} c_n (D^\dagger D)^n\, ,
\end{equation}
or equivalently,
\begin{equation}
    {\mathcal L}_{pol}\,=\, \frac{1}{N_k}\sum_{k}|| D_\theta (k) + D^\dagger_\theta (k) \,-\, \sum_{n=1}^{N} c_n (D^\dagger_\theta (k) D_\theta (k))^n ||^2\,.
\end{equation}
The coefficients $c_n$ define the specific algebraic structure of the fermion. In our numerical demonstration of 2D Dirac operators, we truncate this expansion at $N=5$. We note that, in this investigation, $m_0$ is again randomized during training, while the spectrum is evaluated at $m_0 = -1.0$.

The standard GW relation corresponds precisely to the subset $\{c_1=1.0, \, c_{n>1}=0.0\}$. Our objective is to determine whether the network can independently derive this exact subset of coefficients from a completely agnostic starting state.

\subsection{Differentiable Architecture Search and Global Scale}

We implement a Differentiable Architecture Search (DARTS) inspired mechanism \cite{Liu:2018alq}. We decouple the selection of the algebraic structure from its overall coefficient. The network is parameterized by internal logits $\alpha_n$, which dictate the relative selection probability of each polynomial term via a Softmax function with temperature annealing. Concurrently, a completely independent parameter, $s_{global}$, governs the overall coefficient of the relation.

The coefficients are dynamically determined during training as
\begin{equation}
    c_n \,=\, s_{global} \times \frac{\exp(\alpha_n/\tau)}{\sum_{j} \exp(\alpha_j/\tau)} \times 4^{-(n-1)}\,.
\end{equation}
The factor $4^{-(n-1)}$ represents a standard basis normalization preventing gradient explosion near the Brillouin zone boundaries, not a structural constraint.

Crucially, the algebraic coefficients are initialized without bias toward the standard GW relation. For instance, all structural logits are set to zero ($\alpha_n=0$, ensuring a completely flat probability distribution), and the global scale is initialized to $s_{global}=0.1$. The network is then completely unaware that the physical target scale is $1.0$, nor does it know which combination of $c_n$ is optimal. The total loss function incorporates the polynomial algebraic condition ${\mathcal L}_{pol}$, the locality penalty $\mathcal{L}_{loc}$ in Eq.~(\ref{eq:loc}) with the decay parameter $\alpha = 0.8$ and the lattice size $L=32$, and the doubler-decoupling conditions $\mathcal{L}_{pin}$ in Eq.~(\ref{eq:pin}) as
\begin{equation}
    {\mathcal L}_{total} \,=\, w_{pol} {\mathcal L}_{pol} \,+\, w_{pin} {\mathcal L}_{pin} \,+\, w_{loc} {\mathcal L}_{loc} \,.
\end{equation}
For this autonomous discovery phase, the loss weights are set to $w_{pol} = 1.0$, $w_{pin} = 0.1$, and $w_{loc} = 0.01$, and the network is trained using the Adam optimizer with a learning rate of $\eta = 0.002$. The Softmax temperature $\tau$ is exponentially annealed from $\tau = 0.5$ down to $\tau = 0.01$ over the 10,000 epochs. The network must optimize its own algebraic rules to satisfy these basic requirements.

\subsection{Results: Emergence of the GW Relation}

\begin{figure}[t]
\centering
\includegraphics[width=0.5\linewidth]{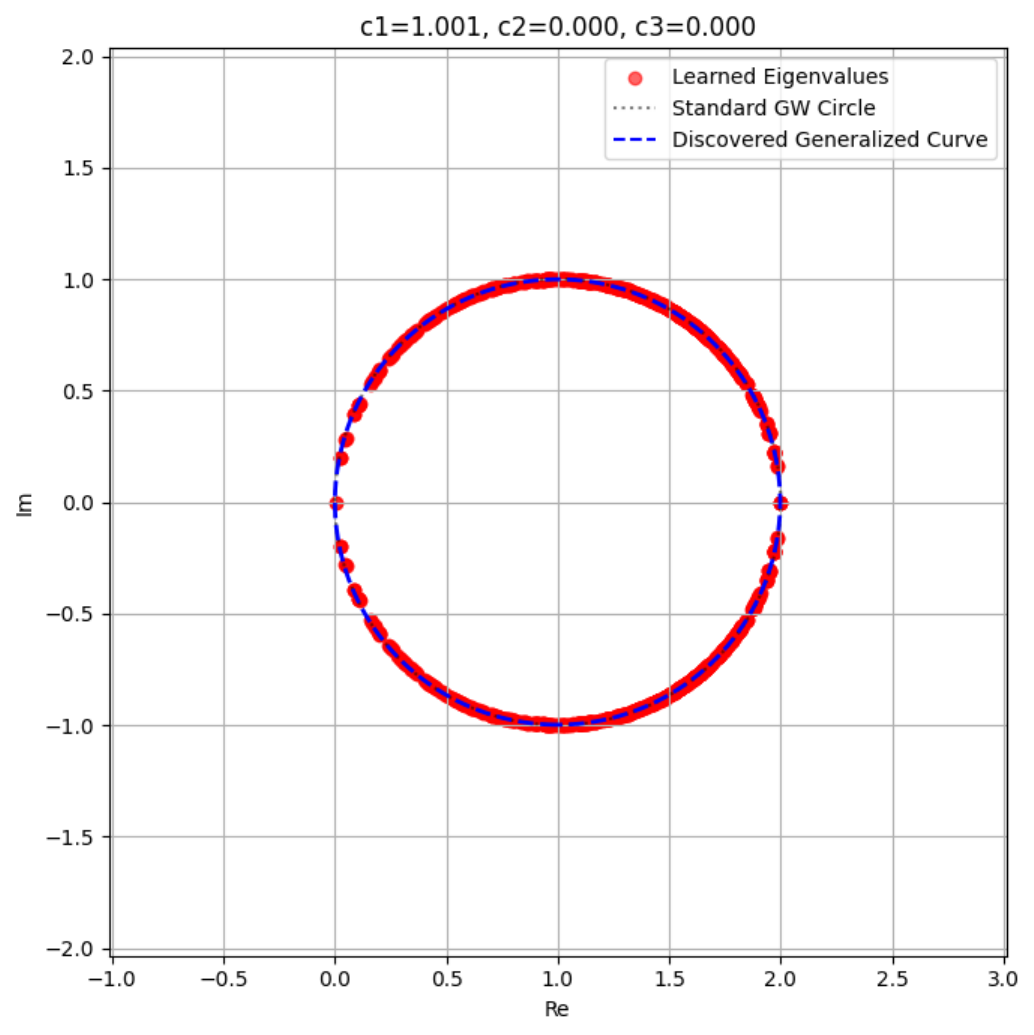}
\caption{The learned eigenvalue spectrum in 2D. Guided by the generalized polynomial constraint, spatial locality, and doubler-decoupling conditions, the network selects $s_{global} \to 1.001\, (c_{1}=1.001)$ and zeroes out higher-order coefficients ($c_{n>1}=0.000$), indicating that the standard GW relation is selected as the preferred solution within the present polynomial optimization setup. The parameters are $\alpha=0.8$, $L=32$ and $w_{pol} = 1.0$, $w_{pin} = 0.1$, $w_{loc} = 0.01$.}
\label{fig:standard_gw_2d}
\end{figure}

The dynamics of the optimization process illustrate how the proposed framework performs machine-assisted algebraic-structure search. Starting from random initial network weights, the initial outputs yield arbitrary, non-local spectra that entirely violate chiral symmetry. However, constrained by the generalized polynomial framework and guided purely by the necessity to fulfill the doubler-pinning condition ($D( k_{doub}) = 2$) and the spatial locality penalty, the network systematically traverses the loss landscape. 

As training progresses over 10,000 epochs, the network smoothly transitions from a broad exploration of the generalized polynomial space to a sharp decision via temperature annealing in the Softmax function ($\tau = 0.5 \to 0.01$). During the early phases, the optimization suppresses higher-order terms, suggesting that they are disfavored
by the combined polynomial, locality, and pinning losses in the present setup. Consequently, the Softmax gating dynamically suppresses the higher-order logits, completely isolating the $n=1$ structural term ($c_{n>1} \to 0.000$). Simultaneously, the network incrementally drives the global scale from its naive starting point of $0.1$ up to $\sim 1.00$. As shown in Fig.~\ref{fig:standard_gw_2d}, by the end of the 10,000 epochs, the resultant eigenvalue spectrum eventually converges to form a GW circle in the complex plane, confirming the emergence of the GW formulation. The near zero mode in this discovered operator deviates from the origin by only $\sim 4.9084 \times 10^{-3}$.
Let us show the explicit results of coefficients:
\begin{align}
{\rm Initial:}&\,\,\, s_{global}=0.1,\,\, c_{n} = 0.1 \times 0.2  \times 4^{-n+1}\,\,\,\,(\alpha_n = 0)\,,
\nonumber\\
{\rm Final:}&\,\,\, s_{global}=1.001,\,c_{1} = 1.001,\,c_{n>1} \simeq 0.000\,.
\end{align}
The other initial coefficients also result in the same final coefficients:
\begin{align}
{\rm Initial:}&\,\,\, s_{global}=0.1,\, c_{1} = 0.01,\,c_{2}=0.016,\, c_{3}=0.001,\, c_{4,5}\simeq 0.000\,\,\,\,(\alpha_2 = 1.0,\,\alpha_{n\not=2}=0)\,,
\nonumber\\
{\rm Final:}&\,\,\, s_{global}=1.001,\,c_{1} = 1.001,\,c_{n>1} \simeq 0.000\,.
\end{align}
We emphasize that the final 2D Dirac operator learned in this autonomous phase reproduces the overlap operator obtained in Section \ref{sec:overlap_discovery}, with a similar momentum-space profile and real-space exponential locality.

The network was not explicitly tasked with satisfying the Ginsparg-Wilson relation. Instead, starting from a generalized polynomial framework and naive initializations, the optimization selects the $c_1\simeq 1$, $c_{n>1}\simeq 0$ structure as the preferred solution within the prescribed polynomial ansatz and the present loss function.

\subsection{Discovery of the Generalized GW Relation}
The successful autonomous derivation of the standard GW relation raises an intriguing question regarding the structure of the optimization landscape: does the generalized polynomial space contain other physically valid local minima, and can the network autonomously construct them? To investigate this, we maintain the same global scale ($s_{global}=0.1$) but introduce a slight initial bias in the structural logits, starting the network from $[\alpha_1, \alpha_2, \alpha_3, \alpha_4, \alpha_5] = [0.0, 2.0, 0.0, 0.0, 0.0]$. This gently nudges the initial search space toward the $n=2$ term.
 
\begin{figure}[t]
\centering
\includegraphics[width=0.5\linewidth]{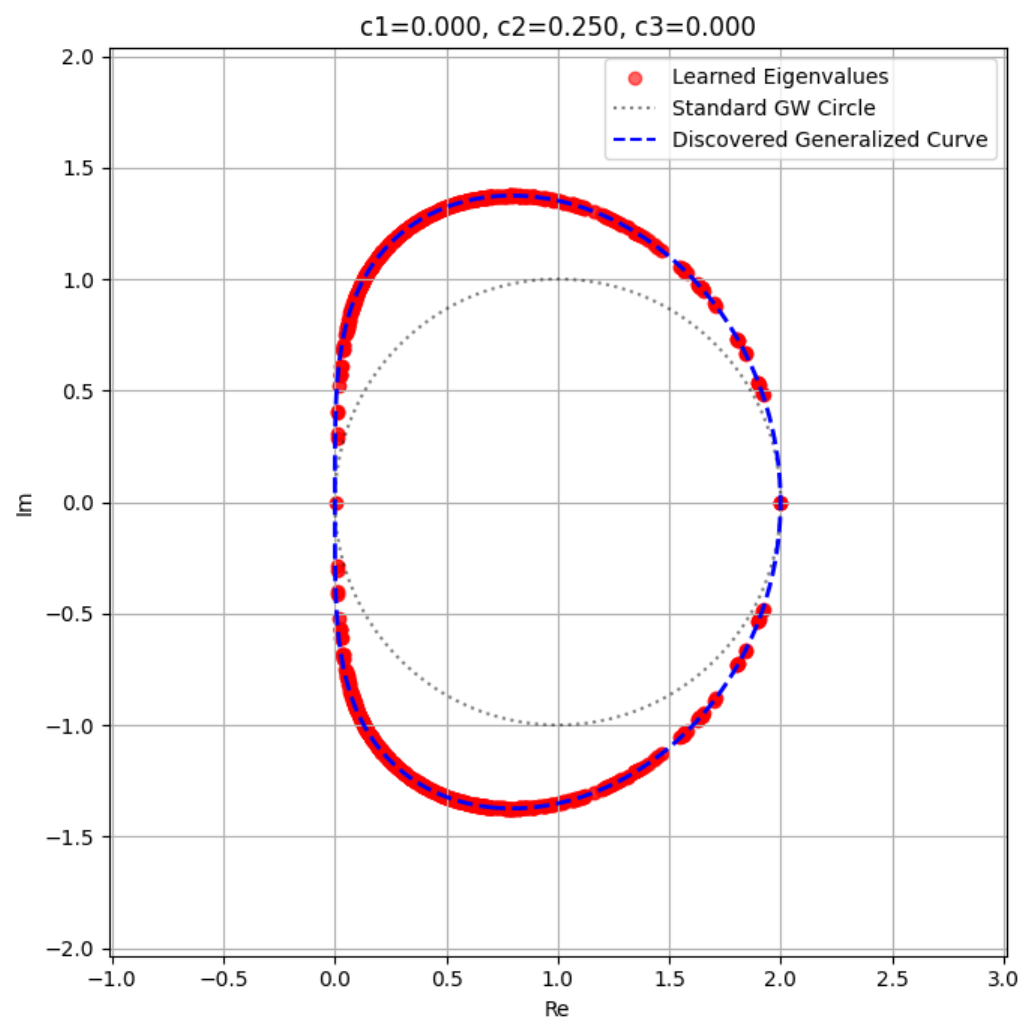} \caption{The learned eigenvalue spectrum corresponding to the autonomous discovery of the generalized Fujikawa relation. By initializing the search with a slight bias toward the $n=2$ term, the network zeroes out all other coefficients and precisely converges to $c_2 = 0.250,\,c_{n\not=2}=0$, showing that the same framework can converge to an alternative higher-order GW-type algebraic structure. The parameters are $\alpha=0.8$, $L=32$ and $w_{pol} = 1.0$, $w_{pin} = 0.1$, $w_{loc} = 0.01$.}
\label{fig:fujikawa_gw_2d}
\end{figure}
 
The resulting training dynamics exhibits a second stable solution in the polynomial search space. As the epochs progress, the network systematically zeroes out the $c_1, c_3, c_4$, and $c_5$ coefficients. Remarkably, by epoch 6,000, the network rigorously drives the global scale to $1.001$ and precisely converges to the coefficient set $\{c_2=0.250, \, c_{n \neq 2}=0.000\}$ as
\begin{align}
{\rm Initial:}&\,\, s_{global}=0.1,\, c_{1} = 0.002,\,c_{2}=0.023,\, c_{3,4,5}\simeq 0\,\,\,\,(\alpha_2 = 2.0,\,\alpha_{n\not=2}=0)\,,
\nonumber\\
{\rm Final:}&\,\, s_{global}=1.001,\,c_{2} = 0.250,\,c_{n\not=2} \simeq 0.000\,.
\end{align}
Mathematically, this specific coefficient selection translates the generalized polynomial ansatz into
\begin{equation}
D \,+\, D^\dagger \,=\, \frac{1}{4} (D^\dagger D)^2\,.
\end{equation}
This is not an arbitrary result; it is precisely the generalized Ginsparg-Wilson relation originally proposed by Fujikawa \cite{Fujikawa:2000my}, which plays a pivotal role in the context of higher-order Dirac operators and index theorems. The corresponding eigenvalue spectrum, depicted in Fig.~\ref{fig:fujikawa_gw_2d}, confirms a symmetric, topologically valid structure distinct from the standard GW circle. Here again, the near zero mode is pinned, with an offset of only $\sim 5.3882 \times 10^{-3}$.

This result suggests that the optimization landscape contains more than the standard GW solution within the chosen polynomial ansatz. Instead, it can access different algebraic structures depending on the initial search bias. Thus, by altering the initial search bias, the same optimization framework can converge to distinct GW-type algebraic structures within the prescribed polynomial ansatz.

\section{\label{sec:conclusion}Conclusion}

We have proposed a novel framework that utilizes Physics-Informed Neural Networks to systematically construct, optimize, and theoretically explore lattice fermions. By integrating doubler-decoupling conditions, spatial locality, and symmetry constraints into the loss function, we treated the formulation of the Dirac operator as a constraint-driven optimization problem. 

We first demonstrated that a neural network can learn the overlap mapping from the Wilson kernel to high numerical accuracy when trained with the GW relation as a soft constraint. 
As a proof of concept, we confirmed that the framework extends to 4D, where the learned spectrum satisfies the GW-circle constraint.
Furthermore, we demonstrated the stability of our framework by analyzing its robustness against variations in the loss weights and its sensitivity to the locality decay parameter.

Most importantly, we elevated this framework from operator construction to machine-assisted algebraic discovery. Within a generalized polynomial ansatz and with an agnostic initial scale, the neural network autonomously suppresses higher-order structural terms and recovers the standard Ginsparg-Wilson relation, including its overall normalization. Furthermore, by exploring different initial search biases, the same framework finds a distinct algebraic structure corresponding to a Fujikawa-type generalized GW relation. These results indicate that the standard and generalized GW relations appear as preferred solutions in the present optimization landscape, selected by the combined requirements of locality, doubler decoupling, and polynomial algebraic consistency.

Looking ahead, there are several promising directions for future research. First, extending this framework to the Hamiltonian formalism will be crucial for tensor network simulations and quantum computing applications. Second, applying this methodology to minimally doubled fermions could systematically identify optimal kernels where specific discrete symmetries such as $\mathcal{C, P, T}$ or hypercubic symmetries are partially relaxed. Third, exploring one-component fermions like staggered fermions under this machine-learning-driven framework may reveal new classes of highly localized operators. Finally, incorporating background gauge fields directly into the training process remains an important next step. These efforts will further establish deep learning as an indispensable theoretical tool for advancing the fundamental mathematical structures of quantum field theory.

The author welcomes discussions on the computational framework and is happy to provide the Python source code used in this study to interested readers upon request.

\begin{acknowledgments}
The author would like to thank Norihiro Tanahashi and Akio Tomiya for encouraging the author to publish the present work. The author also thanks the seminar ``Deep Learning and Physics" and the workshop ``Topology in Lattice Systems 2026" for providing opportunities of useful discussions. This work was supported by Japan Society for the Promotion of Science (JSPS) KAKENHI Grant No.~23K03425 and 22H05118.
\end{acknowledgments}

\appendix

\section{Stability against variations in loss weights}
\label{app:weights}
To verify that the learned operator is not a delicate artifact of a specific choice of hyperparameters, we investigated the stability of the optimization with respect to variations in the loss weights ($w_{GW}$, $w_{pin}$, $w_{loc}$). Specifically, we set $w_{GW} = 1.0$, $w_{pin}=0.1$, and $w_{loc}=0.01$, and performed the same training procedure for $\alpha = 0.8$ and $L=32$ in 2D. 

\begin{figure}[t]
\centering
\includegraphics[width=0.95\linewidth]{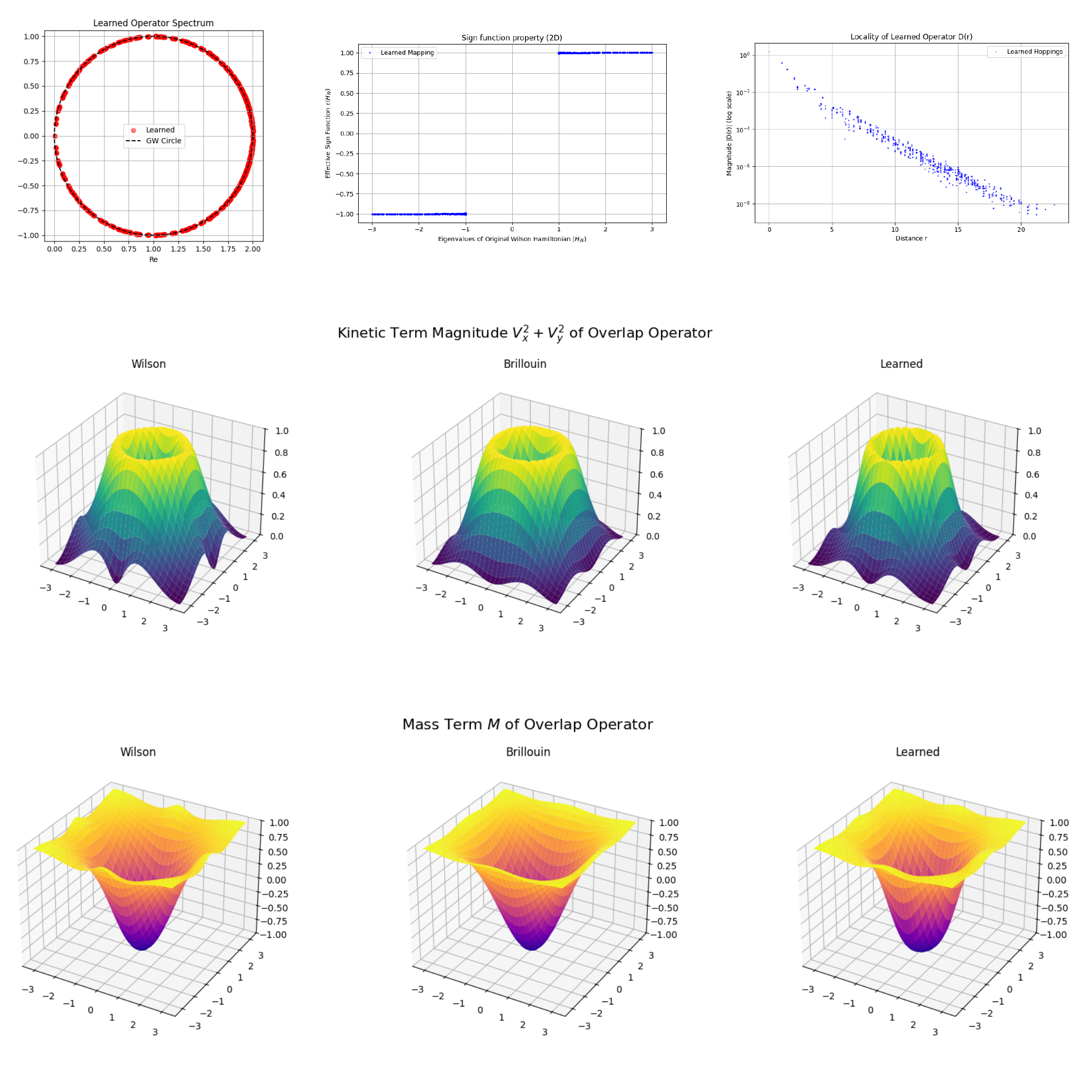}
\caption{Properties of the learned operator trained with a reduced locality penalty ($w_{GW} = 1.0$, $w_{pin} = 0.1$, $w_{loc} = 0.01$) with $\alpha = 0.8$ and $L=32$ in 2D. It seems to converge to an overlap operator, as evidenced by the eigenvalue spectrum and the effective sign function. The dispersion relations exhibit flattened basins near the Brillouin zone boundaries. The optimization of the momentum-space dispersion comes at the cost of real-space locality.}
\label{fig:app_weights}
\end{figure}

As shown in Fig.~\ref{fig:app_weights}, we confirmed that even with this hyperparameter set, the learned operator consistently seems to converge to an overlap operator. Furthermore, the dispersion relation for this set exhibits a relatively flat behavior near the Brillouin zone boundary, yielding a more Brillouin-fermion-like overlap operator. However, due to the reduced locality weight $\mu \sim 0.95$, the spatial locality degrades compared to the baseline case ($w_{GW} = 1.0$, $w_{pin}=0.1$, $w_{loc}=0.1$).

\section{Dependence on the locality decay parameter $\alpha$}
\label{app:alpha}
In contrast to the robustness against the loss weights, we found that the optimization landscape is sensitive to the choice of the locality decay parameter $\alpha$. To investigate this dependence, we systematically varied the exponent from $\alpha=0.5$ to $0.9$ with $w_{GW} = 1.0$, $w_{pin} = 0.1$, $w_{loc} = 0.1$ and $L=32$ fixed in 2D. 

\begin{figure}[t]
\centering
\includegraphics[width=0.95\linewidth]{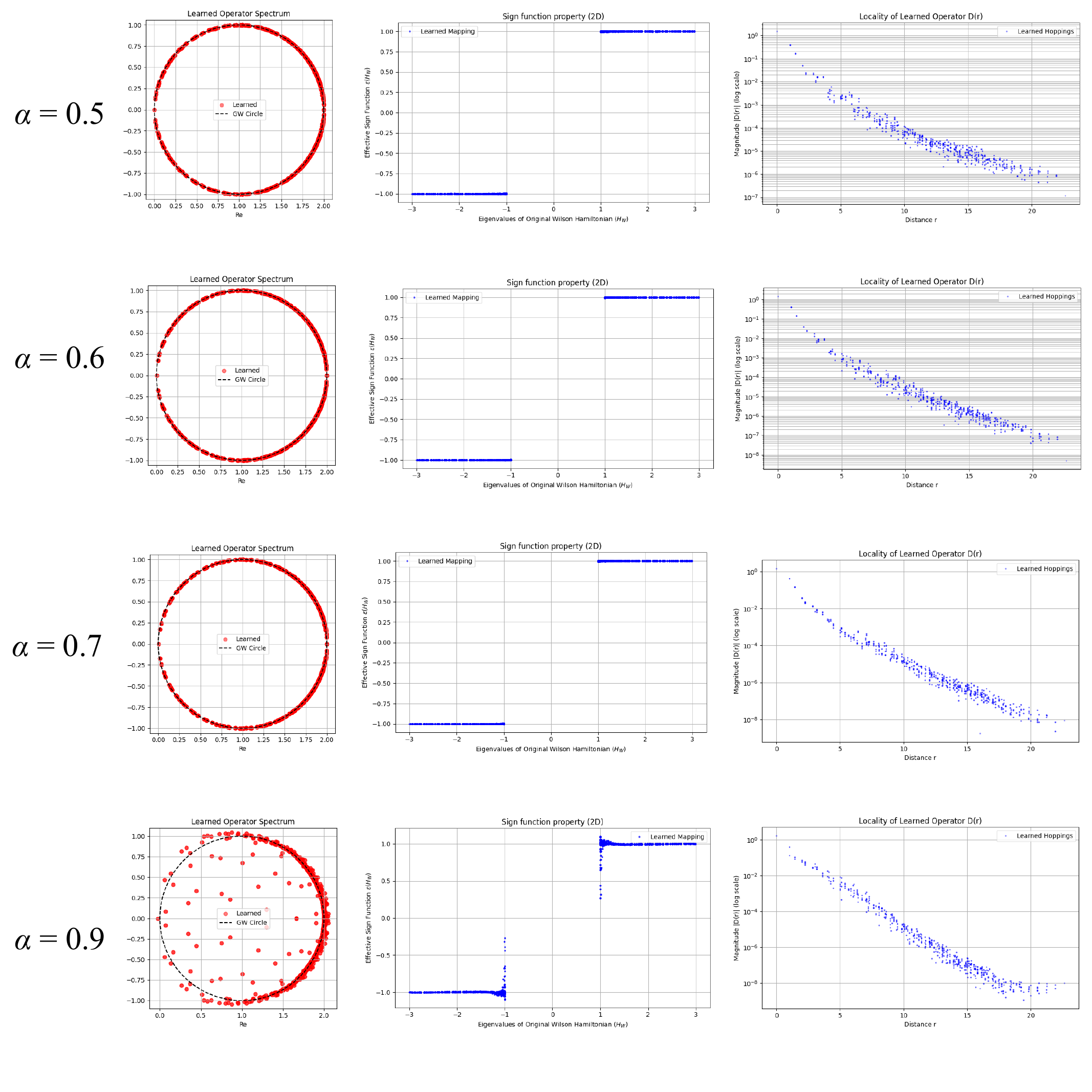}
\caption{The eigenvalue spectra, the effective sign functions and the magnitudes of the real-space hopping parameters $|D(r)|$ for $\alpha=0.5,0.6,0.7,0.9$ with $w_{GW} = 1.0$, $w_{pin} = 0.1$, $w_{loc} = 0.1$ and $L=32$ fixed. For $\alpha = 0.9$, the network fails to reproduce the Ginsparg-Wilson circle of spectrum while it does for $\alpha=0.5,0.6,0.7$.}
\label{fig:app_alpha}
\end{figure}

As illustrated in Fig.~\ref{fig:app_alpha}, when the locality penalty is relatively weak such as $\alpha = 0.5, 0.6, 0.7$, the network seems to successfully converge to the overlap operator, reproducing both the Ginsparg-Wilson circle and the topological sign function. However, the spatial locality naturally degrades as $\alpha$ decreases. 
Conversely, when the penalty is excessively strong such as $\alpha = 0.9$, the network either fails to reproduce the Ginsparg-Wilson relation entirely, or the training process itself becomes highly unstable and fails to converge. This strong sensitivity clearly highlights that a careful adjustment of the real-space locality constraint (around $\alpha\sim 0.8$) is a crucial ingredient for guiding the machine-learning framework to valid chiral lattice fermion formulations.

\bibliographystyle{utphys}
\bibliography{./QFT,./refs}

\providecommand{\href}[2]{#2}\begingroup\raggedright\begin{thebibliography}{10}

\bibitem{Karsten:1980wd}
L.~H. Karsten and J.~Smit, ``{Lattice Fermions: Species Doubling, Chiral
  Invariance, and the Triangle Anomaly},''
  \href{http://dx.doi.org/10.1016/0550-3213(81)90549-6}{{\em Nucl. Phys. B}
  {\bfseries 183} (1981) 103}.

\bibitem{Nielsen:1980rz}
H.~B. Nielsen and M.~Ninomiya, ``{Absence of Neutrinos on a Lattice. 1. Proof
  by Homotopy Theory},''
\href{http://dx.doi.org/10.1016/0550-3213(81)90361-8}{{\em Nucl. Phys.}
  {\bfseries B185} (1981) 20}.

\bibitem{Nielsen:1981xu}
H.~B. Nielsen and M.~Ninomiya, ``{Absence of Neutrinos on a Lattice. 2.
  Intuitive Topological Proof},''
\href{http://dx.doi.org/10.1016/0550-3213(81)90524-1}{{\em Nucl. Phys.}
  {\bfseries B193} (1981) 173--194}.

\bibitem{Nielsen:1981hk}
H.~B. Nielsen and M.~Ninomiya, ``{No Go Theorem for Regularizing Chiral
  Fermions},''
\href{http://dx.doi.org/10.1016/0370-2693(81)91026-1}{{\em Phys. Lett.}
  {\bfseries 105B} (1981) 219--223}.

\bibitem{Wilson:1975id}
K.~G. Wilson, \href{http://dx.doi.org/10.1007/978-1-4613-4208-3_6}{``{Quarks
  and Strings on a Lattice},''} in {\em {New Phenomena in Subnuclear Physics:
  Proceedings, International School of Subnuclear Physics, Erice, Sicily, Jul
  11-Aug 1 1975. Part A}}, pp.~69--142.
\newblock 1975.
\newblock
\url{https://doi.org/10.1007/978-1-4613-4208-3_6}.
\newblock

\bibitem{Kogut:1974ag}
J.~B. Kogut and L.~Susskind, ``{Hamiltonian Formulation of Wilson's Lattice
  Gauge Theories},''
\href{http://dx.doi.org/10.1103/PhysRevD.11.395}{{\em Phys. Rev.} {\bfseries
  D11} (1975) 395--408}.

\bibitem{Susskind:1976jm}
L.~Susskind, ``{Lattice Fermions},''
\href{http://dx.doi.org/10.1103/PhysRevD.16.3031}{{\em Phys. Rev.} {\bfseries
  D16} (1977) 3031--3039}.

\bibitem{Ginsparg:1981bj}
P.~H. Ginsparg and K.~G. Wilson, ``{A Remnant of Chiral Symmetry on the
  Lattice},''
\href{http://dx.doi.org/10.1103/PhysRevD.25.2649}{{\em Phys. Rev.} {\bfseries
  D25} (1982) 2649}.

\bibitem{Luscher:1998pqa}
M.~Luscher, ``{Exact chiral symmetry on the lattice and the Ginsparg-Wilson
  relation},'' \href{http://dx.doi.org/10.1016/S0370-2693(98)00423-7}{{\em
  Phys. Lett. B} {\bfseries 428} (1998) 342--345},
  \href{http://arxiv.org/abs/hep-lat/9802011}{{\ttfamily
  arXiv:hep-lat/9802011}}.

\bibitem{Neuberger:1998wv}
H.~Neuberger, ``{More about exactly massless quarks on the lattice},''
  \href{http://dx.doi.org/10.1016/S0370-2693(98)00355-4}{{\em Phys. Lett.}
  {\bfseries B427} (1998) 353--355},
\href{http://arxiv.org/abs/hep-lat/9801031}{{\ttfamily arXiv:hep-lat/9801031
  [hep-lat]}}.

\bibitem{Kaplan:1992bt}
D.~B. Kaplan, ``{A Method for simulating chiral fermions on the lattice},''
  \href{http://dx.doi.org/10.1016/0370-2693(92)91112-M}{{\em Phys. Lett.}
  {\bfseries B288} (1992) 342--347},
\href{http://arxiv.org/abs/hep-lat/9206013}{{\ttfamily arXiv:hep-lat/9206013
  [hep-lat]}}.

\bibitem{Shamir:1993zy}
Y.~Shamir, ``{Chiral fermions from lattice boundaries},''
  \href{http://dx.doi.org/10.1016/0550-3213(93)90162-I}{{\em Nucl. Phys.}
  {\bfseries B406} (1993) 90--106},
\href{http://arxiv.org/abs/hep-lat/9303005}{{\ttfamily arXiv:hep-lat/9303005
  [hep-lat]}}.

\bibitem{Furman:1994ky}
V.~Furman and Y.~Shamir, ``{Axial symmetries in lattice QCD with Kaplan
  fermions},'' \href{http://dx.doi.org/10.1016/0550-3213(95)00031-M}{{\em Nucl.
  Phys. B} {\bfseries 439} (1995) 54--78},
  \href{http://arxiv.org/abs/hep-lat/9405004}{{\ttfamily
  arXiv:hep-lat/9405004}}.

\bibitem{Chiu:2002eh}
T.-W. Chiu, T.-H. Hsieh, C.-H. Huang, and T.-R. Huang, ``{A Note on the
  Zolotarev optimal rational approximation for the overlap Dirac operator},''
  \href{http://dx.doi.org/10.1103/PhysRevD.66.114502}{{\em Phys. Rev. D}
  {\bfseries 66} (2002) 114502},
  \href{http://arxiv.org/abs/hep-lat/0206007}{{\ttfamily
  arXiv:hep-lat/0206007}}.

\bibitem{Chiu:2002ir}
T.-W. Chiu, ``{Optimal domain wall fermions},''
  \href{http://dx.doi.org/10.1103/PhysRevLett.90.071601}{{\em Phys. Rev. Lett.}
  {\bfseries 90} (2003) 071601},
  \href{http://arxiv.org/abs/hep-lat/0209153}{{\ttfamily
  arXiv:hep-lat/0209153}}.

\bibitem{Gattringer:2000js}
C.~Gattringer, ``{A New approach to Ginsparg-Wilson fermions},''
  \href{http://dx.doi.org/10.1103/PhysRevD.63.114501}{{\em Phys. Rev. D}
  {\bfseries 63} (2001) 114501},
  \href{http://arxiv.org/abs/hep-lat/0003005}{{\ttfamily
  arXiv:hep-lat/0003005}}.

\bibitem{Bietenholz:1998ut}
W.~Bietenholz, ``{Solutions of the Ginsparg-Wilson relation and improved domain
  wall fermions},'' \href{http://dx.doi.org/10.1007/s100520050364}{{\em Eur.
  Phys. J. C} {\bfseries 6} (1999) 537--547},
  \href{http://arxiv.org/abs/hep-lat/9803023}{{\ttfamily
  arXiv:hep-lat/9803023}}.

\bibitem{Bietenholz:1999km}
W.~Bietenholz and I.~Hip, ``{The Scaling of exact and approximate
  Ginsparg-Wilson fermions},''
  \href{http://dx.doi.org/10.1016/S0550-3213(99)00477-0}{{\em Nucl. Phys. B}
  {\bfseries 570} (2000) 423--451},
  \href{http://arxiv.org/abs/hep-lat/9902019}{{\ttfamily
  arXiv:hep-lat/9902019}}.

\bibitem{Durr:2010ch}
S.~D\"urr and G.~Koutsou, ``{Brillouin improvement for Wilson fermions},''
  \href{http://dx.doi.org/10.1103/PhysRevD.83.114512}{{\em Phys. Rev. D}
  {\bfseries 83} (Jun, 2011) 114512},
  \href{http://arxiv.org/abs/1012.3615}{{\ttfamily arXiv:1012.3615 [hep-lat]}}.
  \url{https://link.aps.org/doi/10.1103/PhysRevD.83.114512}.

\bibitem{Durr:2012dw}
S.~D\"urr, G.~Koutsou, and T.~Lippert, ``{Meson and Baryon dispersion relations
  with Brillouin fermions},''
  \href{http://dx.doi.org/10.1103/PhysRevD.86.114514}{{\em Phys. Rev. D}
  {\bfseries 86} (2012) 114514},
  \href{http://arxiv.org/abs/1208.6270}{{\ttfamily arXiv:1208.6270 [hep-lat]}}.

\bibitem{Cho:2013yha}
Y.-G. Cho, S.~Hashimoto, J.-I. Noaki, A.~Juttner, and M.~Marinkovic,
  ``{$O(a^2)$-improved actions for heavy quarks and scaling studies on quenched
  lattices},'' \href{http://dx.doi.org/10.22323/1.187.0255}{{\em PoS}
  {\bfseries LATTICE2013} (2014) 255},
  \href{http://arxiv.org/abs/1312.4630}{{\ttfamily arXiv:1312.4630 [hep-lat]}}.

\bibitem{Cho:2015ffa}
Y.-G. Cho, S.~Hashimoto, A.~Juttner, T.~Kaneko, M.~Marinkovic, J.-I. Noaki, and
  J.~T. Tsang, ``{Improved lattice fermion action for heavy quarks},''
  \href{http://dx.doi.org/10.1007/JHEP05(2015)072}{{\em JHEP} {\bfseries 05}
  (2015) 072}, \href{http://arxiv.org/abs/1504.01630}{{\ttfamily
  arXiv:1504.01630 [hep-lat]}}.

\bibitem{Durr:2017wfi}
S.~D\"urr and G.~Koutsou, ``{On the suitability of the Brillouin action as a
  kernel to the overlap procedure},''
  \href{http://arxiv.org/abs/1701.00726}{{\ttfamily arXiv:1701.00726
  [hep-lat]}}.

\bibitem{Creutz:2010bm}
M.~Creutz, T.~Kimura, and T.~Misumi, ``{Index Theorem and Overlap Formalism
  with Naive and Minimally Doubled Fermions},''
  \href{http://dx.doi.org/10.1007/JHEP12(2010)041}{{\em JHEP} {\bfseries 12}
  (2010) 041},
\href{http://arxiv.org/abs/1011.0761}{{\ttfamily arXiv:1011.0761 [hep-lat]}}.

\bibitem{Misumi:2012eh}
T.~Misumi, ``{New fermion discretizations and their applications},''
  \href{http://dx.doi.org/10.22323/1.164.0005}{{\em PoS} {\bfseries
  LATTICE2012} (2012) 005},
\href{http://arxiv.org/abs/1211.6999}{{\ttfamily arXiv:1211.6999 [hep-lat]}}.

\bibitem{Raissi:2017zsi}
M.~Raissi, P.~Perdikaris, and G.~E. Karniadakis, ``{Physics-informed neural
  networks: A deep learning framework for solving forward and inverse problems
  involving nonlinear partial differential equations},''
  \href{http://dx.doi.org/10.1016/j.jcp.2018.10.045}{{\em J. Comput. Phys.}
  {\bfseries 378} (2019) 686--707},
  \href{http://arxiv.org/abs/1711.10561}{{\ttfamily arXiv:1711.10561 [cs.AI]}}.

\bibitem{Boyda:2022nmh}
D.~Boyda {\em et~al.}, ``{Applications of Machine Learning to Lattice Quantum
  Field Theory},'' in {\em {Snowmass 2021}}.
\newblock 2, 2022.
\newblock \href{http://arxiv.org/abs/2202.05838}{{\ttfamily arXiv:2202.05838
  [hep-lat]}}.

\bibitem{Tomiya:2025quf}
A.~Tomiya, ``{Machine Learning for Lattice QCD},''
  \href{http://dx.doi.org/10.7566/JPSJ.94.031006}{{\em J. Phys. Soc. Jap.}
  {\bfseries 94} no.~3, (2025) 031006}.

\bibitem{Wenger:2026sjp}
U.~Wenger, ``{Machine learning for four-dimensional SU(3) lattice gauge
  theories},'' in {\em {42th International Symposium on Lattice Field Theory}}.
\newblock 4, 2026.
\newblock \href{http://arxiv.org/abs/2604.12416}{{\ttfamily arXiv:2604.12416
  [hep-lat]}}.

\bibitem{Holland:2023ews}
K.~Holland, A.~Ipp, D.~I. M{\"u}ller, and U.~Wenger, ``{Fixed point actions
  from convolutional neural networks},''
  \href{http://dx.doi.org/10.22323/1.453.0038}{{\em PoS} {\bfseries
  LATTICE2023} (2024) 038}, \href{http://arxiv.org/abs/2311.17816}{{\ttfamily
  arXiv:2311.17816 [hep-lat]}}.

\bibitem{Holland:2024muu}
K.~Holland, A.~Ipp, D.~I. M{\"u}ller, and U.~Wenger, ``{Machine learning a
  fixed point action for SU(3) gauge theory with a gauge equivariant
  convolutional neural network},''
  \href{http://dx.doi.org/10.1103/PhysRevD.110.074502}{{\em Phys. Rev. D}
  {\bfseries 110} no.~7, (2024) 074502},
  \href{http://arxiv.org/abs/2401.06481}{{\ttfamily arXiv:2401.06481
  [hep-lat]}}.

\bibitem{Wenger:2025sre}
U.~Wenger, K.~Holland, A.~Ipp, and D.~I. M{\"u}ller, ``{HMC and gradient flow
  with machine-learned classically perfect fixed point actions},''
  \href{http://dx.doi.org/10.22323/1.466.0466}{{\em PoS} {\bfseries
  LATTICE2024} (2025) 466}, \href{http://arxiv.org/abs/2502.03315}{{\ttfamily
  arXiv:2502.03315 [hep-lat]}}.

\bibitem{Holland:2025fsa}
K.~Holland, A.~Ipp, D.~I. M{\"u}ller, and U.~Wenger, ``{Machine-Learned
  Renormalization-Group-Improved Gauge Actions and Classically Perfect Gradient
  Flows},'' \href{http://dx.doi.org/10.1103/k41k-2pnc}{{\em Phys. Rev. Lett.}
  {\bfseries 136} no.~3, (2026) 031901},
  \href{http://arxiv.org/abs/2504.15870}{{\ttfamily arXiv:2504.15870
  [hep-lat]}}.

\bibitem{Yasunaga:2026smj}
S.~Yasunaga, K.~Yoshimura, A.~Tomiya, and Y.~Nagai, ``{Parameter Optimization
  of Domain-Wall Fermion using Machine Learning},'' in {\em {42th International
  Symposium on Lattice Field Theory}}.
\newblock 3, 2026.
\newblock \href{http://arxiv.org/abs/2603.16329}{{\ttfamily arXiv:2603.16329
  [hep-lat]}}.

\bibitem{Liu:2018alq}
H.~Liu, K.~Simonyan, and Y.~Yang, ``{DARTS: Differentiable Architecture
  Search},'' \href{http://arxiv.org/abs/1806.09055}{{\ttfamily arXiv:1806.09055
  [cs.LG]}}.

\bibitem{Fujikawa:2000my}
K.~Fujikawa, ``{Algebraic generalization of the Ginsparg-Wilson relation},''
  \href{http://dx.doi.org/10.1016/S0550-3213(00)00395-3}{{\em Nucl. Phys. B}
  {\bfseries 589} (2000) 487--503},
  \href{http://arxiv.org/abs/hep-lat/0004012}{{\ttfamily
  arXiv:hep-lat/0004012}}.

\bibitem{Clancy:2023ino}
M.~Clancy, D.~B. Kaplan, and H.~Singh, ``{Generalized Ginsparg-Wilson
  relations},'' \href{http://dx.doi.org/10.1103/PhysRevD.109.014502}{{\em Phys.
  Rev. D} {\bfseries 109} no.~1, (2024) 014502},
  \href{http://arxiv.org/abs/2309.08542}{{\ttfamily arXiv:2309.08542
  [hep-lat]}}.

\bibitem{Hasenfratz:1993sp}
P.~Hasenfratz and F.~Niedermayer, ``{Perfect lattice action for asymptotically
  free theories},'' \href{http://dx.doi.org/10.1016/0550-3213(94)90261-5}{{\em
  Nucl. Phys. B} {\bfseries 414} (1994) 785--814},
  \href{http://arxiv.org/abs/hep-lat/9308004}{{\ttfamily
  arXiv:hep-lat/9308004}}.

\bibitem{Bietenholz:1995cy}
W.~Bietenholz and U.~J. Wiese, ``{Perfect lattice actions for quarks and
  gluons},'' \href{http://dx.doi.org/10.1016/0550-3213(95)00678-8}{{\em Nucl.
  Phys. B} {\bfseries 464} (1996) 319--352},
  \href{http://arxiv.org/abs/hep-lat/9510026}{{\ttfamily
  arXiv:hep-lat/9510026}}.

\bibitem{Bietenholz:2006ni}
W.~Bietenholz, ``{Optimised Dirac Operators on the Lattice: Construction,
  Properties and Applications},''
  \href{http://dx.doi.org/10.1002/prop.200710397}{{\em Fortsch. Phys.}
  {\bfseries 56} (2008) 107--180},
  \href{http://arxiv.org/abs/hep-lat/0611030}{{\ttfamily
  arXiv:hep-lat/0611030}}.

\bibitem{Karsten:1981gd}
L.~H. Karsten, ``{Lattice Fermions in Euclidean Space-time},''
  \href{http://dx.doi.org/10.1016/0370-2693(81)90133-7}{{\em Phys. Lett. B}
  {\bfseries 104} (1981) 315--319}.

\bibitem{Wilczek:1987kw}
F.~Wilczek, ``{ON LATTICE FERMIONS},''
  \href{http://dx.doi.org/10.1103/PhysRevLett.59.2397}{{\em Phys. Rev. Lett.}
  {\bfseries 59} (1987) 2397}.

\bibitem{Creutz:2007af}
M.~Creutz, ``{Four-dimensional graphene and chiral fermions},''
  \href{http://dx.doi.org/10.1088/1126-6708/2008/04/017}{{\em JHEP} {\bfseries
  04} (2008) 017}, \href{http://arxiv.org/abs/0712.1201}{{\ttfamily
  arXiv:0712.1201 [hep-lat]}}.

\bibitem{Borici:2007kz}
A.~Borici, ``{Creutz fermions on an orthogonal lattice},''
  \href{http://dx.doi.org/10.1103/PhysRevD.78.074504}{{\em Phys. Rev. D}
  {\bfseries 78} (2008) 074504},
  \href{http://arxiv.org/abs/0712.4401}{{\ttfamily arXiv:0712.4401 [hep-lat]}}.

\bibitem{Bedaque:2008xs}
P.~F. Bedaque, M.~I. Buchoff, B.~C. Tiburzi, and A.~Walker-Loud, ``{Broken
  Symmetries from Minimally Doubled Fermions},''
  \href{http://dx.doi.org/10.1016/j.physletb.2008.03.034}{{\em Phys. Lett. B}
  {\bfseries 662} (2008) 449--455},
  \href{http://arxiv.org/abs/0801.3361}{{\ttfamily arXiv:0801.3361 [hep-lat]}}.

\bibitem{Bedaque:2008jm}
P.~F. Bedaque, M.~I. Buchoff, B.~C. Tiburzi, and A.~Walker-Loud, ``{Search for
  Fermion Actions on Hyperdiamond Lattices},''
  \href{http://dx.doi.org/10.1103/PhysRevD.78.017502}{{\em Phys. Rev. D}
  {\bfseries 78} (2008) 017502},
  \href{http://arxiv.org/abs/0804.1145}{{\ttfamily arXiv:0804.1145 [hep-lat]}}.

\bibitem{Capitani:2009yn}
S.~Capitani, J.~Weber, and H.~Wittig, ``{Minimally doubled fermions at one
  loop},'' \href{http://dx.doi.org/10.1016/j.physletb.2009.09.050}{{\em Phys.
  Lett. B} {\bfseries 681} (2009) 105--112},
  \href{http://arxiv.org/abs/0907.2825}{{\ttfamily arXiv:0907.2825 [hep-lat]}}.

\bibitem{Kimura:2009qe}
T.~Kimura and T.~Misumi, ``{Characters of Lattice Fermions Based on the
  Hyperdiamond Lattice},'' \href{http://dx.doi.org/10.1143/PTP.124.415}{{\em
  Prog. Theor. Phys.} {\bfseries 124} (2010) 415--432},
  \href{http://arxiv.org/abs/0907.1371}{{\ttfamily arXiv:0907.1371 [hep-lat]}}.

\bibitem{Kimura:2009di}
T.~Kimura and T.~Misumi, ``{Lattice Fermions Based on Higher-Dimensional
  Hyperdiamond Lattices},'' \href{http://dx.doi.org/10.1143/PTP.123.63}{{\em
  Prog. Theor. Phys.} {\bfseries 123} (2010) 63--78},
  \href{http://arxiv.org/abs/0907.3774}{{\ttfamily arXiv:0907.3774 [hep-lat]}}.

\bibitem{Creutz:2010cz}
M.~Creutz and T.~Misumi, ``{Classification of Minimally Doubled Fermions},''
  \href{http://dx.doi.org/10.1103/PhysRevD.82.074502}{{\em Phys. Rev. D}
  {\bfseries 82} (2010) 074502},
  \href{http://arxiv.org/abs/1007.3328}{{\ttfamily arXiv:1007.3328 [hep-lat]}}.

\bibitem{Creutz:2010qm}
M.~Creutz, ``{Minimal doubling and point splitting},''
  \href{http://dx.doi.org/10.22323/1.105.0078}{{\em PoS} {\bfseries
  LATTICE2010} (2010) 078}, \href{http://arxiv.org/abs/1009.3154}{{\ttfamily
  arXiv:1009.3154 [hep-lat]}}.

\bibitem{Capitani:2010nn}
S.~Capitani, M.~Creutz, J.~Weber, and H.~Wittig, ``{Renormalization of
  minimally doubled fermions},''
  \href{http://dx.doi.org/10.1007/JHEP09(2010)027}{{\em JHEP} {\bfseries 09}
  (2010) 027}, \href{http://arxiv.org/abs/1006.2009}{{\ttfamily arXiv:1006.2009
  [hep-lat]}}.

\bibitem{Tiburzi:2010bm}
B.~C. Tiburzi, ``{Chiral Lattice Fermions, Minimal Doubling, and the Axial
  Anomaly},'' \href{http://dx.doi.org/10.1103/PhysRevD.82.034511}{{\em Phys.
  Rev. D} {\bfseries 82} (2010) 034511},
  \href{http://arxiv.org/abs/1006.0172}{{\ttfamily arXiv:1006.0172 [hep-lat]}}.

\bibitem{Kamata:2011jn}
S.~Kamata and H.~Tanaka, ``{Minimal Doubling Fermion and Hermiticity},''
  \href{http://dx.doi.org/10.1093/ptep/pts093}{{\em PTEP} {\bfseries 2013}
  (2013) 023B05}, \href{http://arxiv.org/abs/1111.4536}{{\ttfamily
  arXiv:1111.4536 [hep-lat]}}.

\bibitem{Misumi:2012uu}
T.~Misumi, ``{Phase structure for lattice fermions with flavored chemical
  potential terms},'' \href{http://dx.doi.org/10.1007/JHEP08(2012)068}{{\em
  JHEP} {\bfseries 08} (2012) 068},
  \href{http://arxiv.org/abs/1206.0969}{{\ttfamily arXiv:1206.0969 [hep-lat]}}.

\bibitem{Misumi:2012ky}
T.~Misumi, T.~Kimura, and A.~Ohnishi, ``{QCD phase diagram with 2-flavor
  lattice fermion formulations},''
  \href{http://dx.doi.org/10.1103/PhysRevD.86.094505}{{\em Phys. Rev. D}
  {\bfseries 86} (2012) 094505},
  \href{http://arxiv.org/abs/1206.1977}{{\ttfamily arXiv:1206.1977 [hep-lat]}}.

\bibitem{Capitani:2013zta}
S.~Capitani, ``{Reducing the number of counterterms with new minimally doubled
  actions},'' \href{http://dx.doi.org/10.1103/PhysRevD.89.014501}{{\em Phys.
  Rev. D} {\bfseries 89} no.~1, (2014) 014501},
  \href{http://arxiv.org/abs/1307.7497}{{\ttfamily arXiv:1307.7497 [hep-lat]}}.

\bibitem{Capitani:2013iha}
S.~Capitani, ``{New chiral lattice actions of the Borici-Creutz type},''
  \href{http://dx.doi.org/10.1103/PhysRevD.89.074508}{{\em Phys. Rev. D}
  {\bfseries 89} no.~7, (2014) 074508},
  \href{http://arxiv.org/abs/1311.5664}{{\ttfamily arXiv:1311.5664 [hep-lat]}}.

\bibitem{Misumi:2013maa}
T.~Misumi, ``{Fermion Actions extracted from Lattice Super Yang-Mills
  Theories},'' \href{http://dx.doi.org/10.1007/JHEP12(2013)063}{{\em JHEP}
  {\bfseries 12} (2013) 063}, \href{http://arxiv.org/abs/1311.4365}{{\ttfamily
  arXiv:1311.4365 [hep-lat]}}.

\bibitem{Weber:2013tfa}
J.~H. Weber, S.~Capitani, and H.~Wittig, ``{Numerical studies of Minimally
  Doubled Fermions},'' \href{http://dx.doi.org/10.22323/1.187.0122}{{\em PoS}
  {\bfseries LATTICE2013} (2014) 122},
  \href{http://arxiv.org/abs/1312.0488}{{\ttfamily arXiv:1312.0488 [hep-lat]}}.

\bibitem{Weber:2017eds}
J.~H. Weber, {\em {Properties of minimally doubled fermions}}.
\newblock PhD thesis, Mainz U., 2015.
\newblock \href{http://arxiv.org/abs/1706.07104}{{\ttfamily arXiv:1706.07104
  [hep-lat]}}.

\bibitem{Durr:2020yqa}
S.~D\"urr and J.~H. Weber, ``{Dispersion relation and spectral range of
  Karsten-Wilczek and Borici-Creutz fermions},''
  \href{http://dx.doi.org/10.1103/PhysRevD.102.014516}{{\em Phys. Rev. D}
  {\bfseries 102} (Jul, 2020) 014516},
  \href{http://arxiv.org/abs/2003.10803}{{\ttfamily arXiv:2003.10803
  [hep-lat]}}. \url{https://link.aps.org/doi/10.1103/PhysRevD.102.014516}.

\bibitem{Li:2024dpq}
Y.-Y. Li, J.~Wang, and Y.-Z. You, ``{Quantum Many-Body Lattice C-R-T Symmetry:
  Fractionalization, Anomaly, and Symmetric Mass Generation},''
  \href{http://arxiv.org/abs/2412.19691}{{\ttfamily arXiv:2412.19691
  [cond-mat.str-el]}}.

\bibitem{Chatterjee:2024gje}
A.~Chatterjee, S.~D. Pace, and S.-H. Shao, ``{Quantized Axial Charge of
  Staggered Fermions and the Chiral Anomaly},''
  \href{http://dx.doi.org/10.1103/PhysRevLett.134.021601}{{\em Phys. Rev.
  Lett.} {\bfseries 134} no.~2, (2025) 021601},
  \href{http://arxiv.org/abs/2409.12220}{{\ttfamily arXiv:2409.12220
  [hep-th]}}.

\bibitem{Catterall:2025vrx}
S.~Catterall and A.~Pradhan, ``{Symmetries and Anomalies of Hamiltonian
  Staggered Fermions},'' \href{http://arxiv.org/abs/2501.10862}{{\ttfamily
  arXiv:2501.10862 [hep-lat]}}.

\bibitem{Yamaoka:2025sdm}
T.~Yamaoka, ``{Quantized axial charge in the Hamiltonian approach to Wilson
  fermions},'' \href{http://dx.doi.org/10.1007/JHEP10(2025)102}{{\em JHEP}
  {\bfseries 10} (2025) 102}, \href{http://arxiv.org/abs/2504.10263}{{\ttfamily
  arXiv:2504.10263 [hep-lat]}}.

\bibitem{Onogi:2025xir}
T.~Onogi and T.~Yamaoka, ``{Non-singlet conserved charges and anomalies in 3+1
  D staggered fermions},'' \href{http://arxiv.org/abs/2509.04906}{{\ttfamily
  arXiv:2509.04906 [hep-lat]}}.

\bibitem{Aoki:2025vtp}
S.~Aoki, Y.~Kikukawa, and T.~Takemoto, ``{Chiral Anomaly of Kogut-Susskind
  Fermion in (3+1)-dimensional Hamiltonian formalism},''
  \href{http://arxiv.org/abs/2511.06198}{{\ttfamily arXiv:2511.06198
  [hep-lat]}}.

\bibitem{Seiberg:2026icc}
N.~Seiberg and W.~Zhang, ``{Tori, Klein Bottles, and Modulo 8
  Parity/Time-reversal Anomalies of 2+1d Staggered Fermions},''
  \href{http://arxiv.org/abs/2601.01191}{{\ttfamily arXiv:2601.01191
  [hep-th]}}.

\bibitem{Misumi:2025yjf}
T.~Misumi, ``{Minimal-doubling and single-Weyl fermion Hamiltonians},''
  \href{http://dx.doi.org/10.1103/jdy1-qw2s}{{\em Phys. Rev. D} {\bfseries 113}
  no.~7, (2026) 074521}, \href{http://arxiv.org/abs/2512.22609}{{\ttfamily
  arXiv:2512.22609 [hep-lat]}}.

\bibitem{Misumi:2026ckr}
T.~Misumi, T.~Onogi, and T.~Yamaoka, ``{Taste-splitting mass and edge modes in
  $3+1${\textasciitilde}D staggered fermions},''
  \href{http://arxiv.org/abs/2604.02078}{{\ttfamily arXiv:2604.02078
  [hep-lat]}}.

\bibitem{Hernandez:1998et}
P.~Hernandez, K.~Jansen, and M.~Luscher, ``{Locality properties of Neuberger's
  lattice Dirac operator},''
  \href{http://dx.doi.org/10.1016/S0550-3213(99)00213-8}{{\em Nucl. Phys. B}
  {\bfseries 552} (1999) 363--378},
  \href{http://arxiv.org/abs/hep-lat/9808010}{{\ttfamily
  arXiv:hep-lat/9808010}}.

\bibitem{Horvath:1998cm}
I.~Horvath, ``{Ginsparg-Wilson relation and ultralocality},''
  \href{http://dx.doi.org/10.1103/PhysRevLett.81.4063}{{\em Phys. Rev. Lett.}
  {\bfseries 81} (1998) 4063--4066},
  \href{http://arxiv.org/abs/hep-lat/9808002}{{\ttfamily
  arXiv:hep-lat/9808002}}.

\bibitem{Bietenholz:1999dg}
W.~Bietenholz, ``{On the absence of ultralocal Ginsparg-Wilson fermions},''
  \href{http://arxiv.org/abs/hep-lat/9901005}{{\ttfamily
  arXiv:hep-lat/9901005}}.

\end{thebibliography}\endgroup

\end{document}